\documentclass{aa} 

\usepackage{graphicx}
\usepackage{tabularx}
\usepackage{placeins}
\usepackage[rightcaption]{sidecap}
\usepackage[english]{babel}
\usepackage{array}
\usepackage{capt-of}
\usepackage[normalem]{ulem}
\usepackage{pdflscape}

\newcolumntype{L}[1]{>{\raggedright\let\newline\\\arraybackslash\hspace{0pt}}m{#1}}
\newcolumntype{C}[1]{>{\centering\let\newline\\\arraybackslash\hspace{0pt}}m{#1}}
\newcolumntype{R}[1]{>{\raggedleft\let\newline\\\arraybackslash\hspace{0pt}}m{#1}}

\usepackage{txfonts}
\usepackage[colorlinks=true, allcolors=blue]{hyperref}
\def\kms{km~s$^{-1}$}

\graphicspath{{folder/}{Figures/}}

\begin{document} 

   \title{Importance of source structure on complex organics emission}

   \subtitle{I. Observations of CH$_3$OH from low-mass to high-mass protostars}

   \author{
          M. L. van Gelder\inst{1}
          \and
          P. Nazari\inst{1}
          \and
          B. Tabone\inst{1}
          \and
          A. Ahmadi\inst{1}
          \and
          E. F. van Dishoeck\inst{1,2}
          \and 
          M. T. Beltr\'an\inst{3}
          \and
          G. A. Fuller\inst{4,5}
          \and\\
          N. Sakai\inst{6}
          \and
          \'A. S\'anchez-Monge\inst{5}
          \and
          P. Schilke\inst{5}
          \and
          Y.-L. Yang\inst{6,7}
          \and
          Y. Zhang\inst{6}
          }
\institute{
         Leiden Observatory, Leiden University, PO Box 9513, 2300RA Leiden, The Netherlands \\
         \email{vgelder@strw.leidenuniv.nl}
         \and
         Max Planck Institut f\"ur Extraterrestrische Physik (MPE), Giessenbachstrasse 1, 85748 Garching, Germany
         \and
         INAF-Osservatorio Astrofisico di Arcetri, Largo E. Fermi 5, 50125 Firenze, Italy
         \and
         Jodrell Bank Centre for Astrophysics, Department of Physics and Astronomy, University of Manchester, Oxford Road, Manchester, M13 9PL, UK
         \and
         I. Physikalisches Institut, Universit\"at zu K\"oln, Z\"ulpicher Str.77, 50937, K\"oln, Germany
         \and
         RIKEN Cluster for Pioneering Research, Wako-shi, Saitama, 351-0106, Japan
         \and
         Department of Astronomy, University of Virginia, Charlottesville, VA 22904-4235, USA
          }

   \date{Received XXX; accepted XXX}
 
  \abstract
   {Complex organic molecules (COMs) are often observed toward embedded Class 0 and I protostars. However, not all Class 0 and I protostars exhibit COMs emission. }
   {The aim is to study variations in methanol (CH$_3$OH) emission and use this as an observational tracer of hot cores to test if absence of CH$_3$OH emission can be linked to source properties.}
   {A sample of 148 low-mass and high-mass protostars is investigated using new and archival observations  with the Atacama Large Millimeter/submillimeter Array (ALMA) that contain lines of CH$_3$OH and its isotopologues.
   Data for an additional 36 sources are added from the literature giving a total of 184 different sources.
	The warm ($T\gtrsim100$~K) gaseous CH$_3$OH mass, $M_\mathrm{CH_3OH}$, is determined for each source using primarily optically thin isotopologues and compared to a simple toy model of a spherically symmetric infalling envelope that is passively heated the central protostar.}
   {A scatter of more than four orders of magnitude is found for $M_\mathrm{CH_3OH}$ among the low-mass protostars with values ranging between $10^{-7}$~M$_\odot$ and $\lesssim10^{-11}$~M$_\odot$. On average, Class I protostellar systems seem to have less warm $M_\mathrm{CH_3OH}$ ($\lesssim10^{-10}$~M$_\odot$) than younger Class 0 sources ($\sim10^{-7}$~M$_\odot$). High-mass sources in our sample show higher warm $M_\mathrm{CH_3OH}$ up to $\sim 10^{-7}-10^{-3}$~M$_\odot$. 
   To take into account the effect of the source's overall mass on $M_\mathrm{CH_3OH}$, a normalized CH$_3$OH mass is defined as $M_\mathrm{CH_3OH}/M_\mathrm{dust,0}$, where $M_\mathrm{dust,0}$ is the cold + warm dust mass in the disk and inner envelope within a fixed radius measured from the ALMA dust continuum. A correlation between  $M_\mathrm{CH_3OH}/M_\mathrm{dust,0}$ and $L_\mathrm{bol}$ is found.
   Excluding upper limits, a simple power-law fit to the normalized warm CH$_3$OH masses results in $M_\mathrm{CH_3OH}/M_\mathrm{dust,0}\propto L_\mathrm{bol}^{0.70\pm0.05}$ over an $L_\mathrm{bol}$ range of $10^{-1}-10^{6}$~L$_\odot$.
	This is in good agreement with the toy model which predicts that the normalized $M_\mathrm{CH_3OH}$ increases with $L_\mathrm{bol}^{0.75}$ due to the snowline moving outward.
	Sources for which the size of the disk is equivalent or smaller than the estimated 100~K radius fall within the 3$\sigma$ range of the best-fit power-law model, whereas sources with significantly larger disks show up to two orders of magnitude lower normalized warm CH$_3$OH masses.
	}
   {The agreement between sources that are rich in CH$_3$OH with the toy model of a spherically symmetric infalling envelope implies that the thermal structure of the envelopes in these sources is likely not strongly affected by a disk. However, based on the disagreement between the toy model and sources showing less warm CH$_3$OH mass, we suggest that source structure such as a disk can result in colder gas and thus fewer COMs in the gas phase. Additionally, optically thick dust can hide the emission of COMs. Advanced modeling is necessary to quantify the effects of a disk and/or continuum optical depth on the presence of gaseous COMs in young protostellar systems. }


   \maketitle


\section{Introduction}
\label{sec:introduction}
Complex organic molecules (COMs), molecules with six or more atoms \citep[see][for reviews]{Herbst2009,Jorgensen2020}, are commonly observed toward both low-mass and high-mass young embedded protostellar systems \citep[e.g.,][]{Belloche2013,Ceccarelli2014,Jorgensen2016, Bogelund2019,vanGelder2020,Nazari2021,Yang2021}. They are also detected in protostellar outflows \citep[e.g.,][]{Arce2008,Codella2017,Tychoniec2021} and in low abundances in cold clouds \citep{Bacmann2012,Scibelli2021,Jimenez-serra2021}. The simplest COMs such as methanol (CH$_3$OH) and methyl cyanide (CH$_3$CN) have also been detected toward more evolved low-mass Class II systems \citep{Walsh2014,Oberg2015}, in some cases pointing at inheritance between prestellar cores and protoplanetary disks \citep{Booth2021,vanderMarel2021}. It is in particular important to assess the chemical complexity during the embedded Class 0/I phases of star formation since planet formation is thought to start early \citep[e.g.,][]{Harsono2018,Manara2018,Alves2020,Tychoniec2020}.

The earliest Class 0 and I phases are also most suitable for studying COMs in the gas phase since the temperatures in the inner hot cores are high enough to sublimate COMs from the ices \citep[e.g.,][]{vantHoff2020}. However, not all Class~0 and I sources show emission from COMs even when observed at high sensitivity with interferometers like the Atacama Large Millimeter/submillimeter Array (ALMA) and Northern Extended Millimeter Array (NOEMA). \citet{Yang2021} find that CH$_3$OH, the most abundant COM, is present in only 56~\% of the 50 surveyed embedded sources in Perseus. Similarly, \citet{Belloche2020} detect emission from CH$_3$OH toward 50~\% of the 26 low-mass protostars in their sample. A main question remains as to why some embedded sources do show emission of COMs and why others do not. 

One possible explanation for the absence of gaseous COMs emission could be the presence of a disk that casts a shadow on the inner envelope and therefore decreases the temperature throughout the system \citep{Persson2016,Murillo2015,Murillo2018}. Most COMs will then likely be frozen out in the inner envelopes and cold midplanes of these disks. For low-mass sources in the Class~0 phase, these disks are usually small \citep[$<10-50$~au;][]{Segura-Cox2018,Maury2019}, but larger disks up to $\sim100$~au have also been observed \citep[e.g.,][]{Tobin2012,Murillo2013,Sakai2019}. Around more evolved Class I systems, larger disks are more common \citep[e.g.,][]{Harsono2014,Yen2017,ArturdelaVillarmois2019}, although some Class~I sources still also show small $<10$~au disks \citep{Segura-Cox2018}. Furthermore, disk-like structures seem to be present around some high-mass protostars \citep[e.g.,][]{Sanchez-Monge2013,Johnston2015,Ilee2016,Maud2019,Zhang2019,Moscadelli2021}, but not all \citep{Beltran2016}. The effect of a disk on the presence of COMs emission, however, remains largely unknown.

Alternatively, optically thick dust at (sub)millimeter wavelengths can hide the emission of COMs in protostellar systems. This was recently shown to be the case in NGC~1333~IRAS4A1 where \citet{deSimone2020} used the Very Large Array (VLA) to show that this source, which was thought to be poor in COMs based on observations with ALMA \citep{Lopez-Sepulcre2017}, has emission of CH$_3$OH at centimeter wavelengths. Similarly, \citet{Rivilla2017} found high dust opacities affecting the molecular lines of COMs toward the high-mass source G31.41+0.31.

Some sources may also be intrinsically deficient in COMs, although this is not expected based on the large columns of CH$_3$OH ice observed toward prestellar and starless cores \citep[e.g.,][]{Chu2020,Goto2021} and many protostellar systems \citep{Pontoppidan2003,Boogert2008,Oberg2011,Bottinelli2010,Perotti2020,Perotti2021}. Furthermore, recent chemical models of the collapse from prestellar core to protostellar system do not show significant destruction of COMs such as CH$_3$OH \citep{Drozdovskaya2014,Aikawa2020}.

In this work, the amount of warm CH$_3$OH is determined for a large sample of in total 148 sources, covering both low-mass and high-mass protostellar systems. Furthermore, literature values of an additional 36 sources are collected, giving a total sample of 184 sources. The goal is to determine from an observational point of view how source structure such as a disk or optically thick dust affects gaseous COMs emission. Optically thin isotopologues of CH$_3$OH are used to mitigate line optical depth effects. The reduction of the ALMA datasets used in this work and the determination of the amount of warm CH$_3$OH from the ALMA observations are described in Sect.~\ref{sec:methodology}. The results are presented in Sect.~\ref{sec:results} and discussed in Sect.~\ref{sec:discussion}. Our main conclusions are summarized in Sect.~\ref{sec:conclusion}. In this paper, the amount of warm CH$_3$OH is only compared to a simple analytic toy model of a spherically symmetric infalling envelope passively heated by the luminosity of the protostar; a full comparison to sophisticated disk + envelope models including dust optical depth effects is done in a companion paper \citep{Nazari2022}.

\section{Methodology}
\label{sec:methodology}
\subsection{Observations and archival data}
\label{subsec:observations}
The data analyzed in this work consist of a combination of multiple ALMA datasets encompassing 148 different sources. These datasets are all taken in Band~6 ($\sim1.2$~mm) but in different setups covering different transitions of CH$_3$OH and its $^{13}$C and $^{18}$O isotopologues. Moreover, the observations are taken with different beam sizes and sensitivities and the observed sources are located at various distances which need to be taken into account in the analysis (see Sect.~\ref{subsec:calc_warm_methanol_mass}). The full list of sources and their observational properties are presented in Table~\ref{tab:observations}.

The reduction of the 2017.1.01174.S (PI: E.F. van Dishoeck) dataset, containing four protostars in Perseus and three in Serpens, is described by \cite{vanGelder2020}. Only the Band~6 data with a $0.45^{\prime\prime}$ beam are used in this work, which contain transitions of both $^{13}$C and $^{18}$O isotopologues of CH$_3$OH. Three out of seven sources in this program show emission from COMs. For Serpens SMM3, archival pipeline imaged product data are used from the 2017.1.01350.S (PI: {\L}. Tychoniec) program \citep{Tychoniec2021}. 

\begin{figure*}[]
\includegraphics[width=0.33\linewidth]{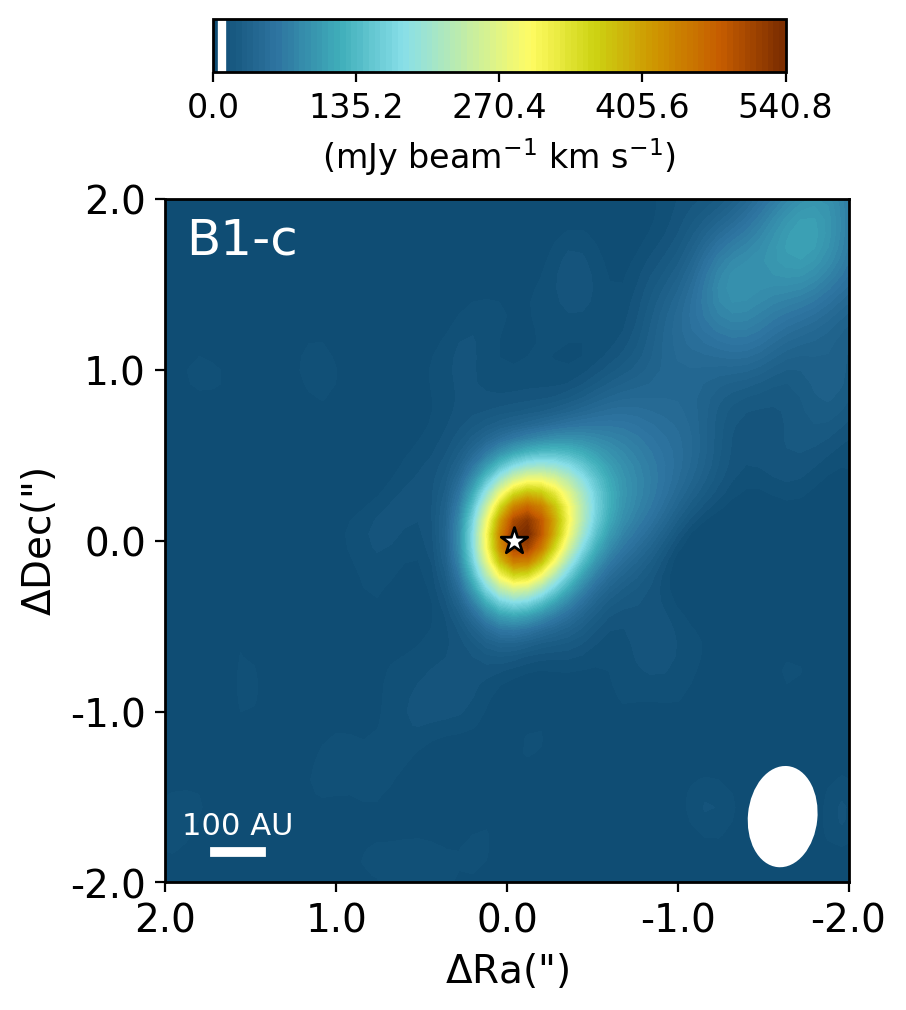}
\includegraphics[width=0.33\linewidth]{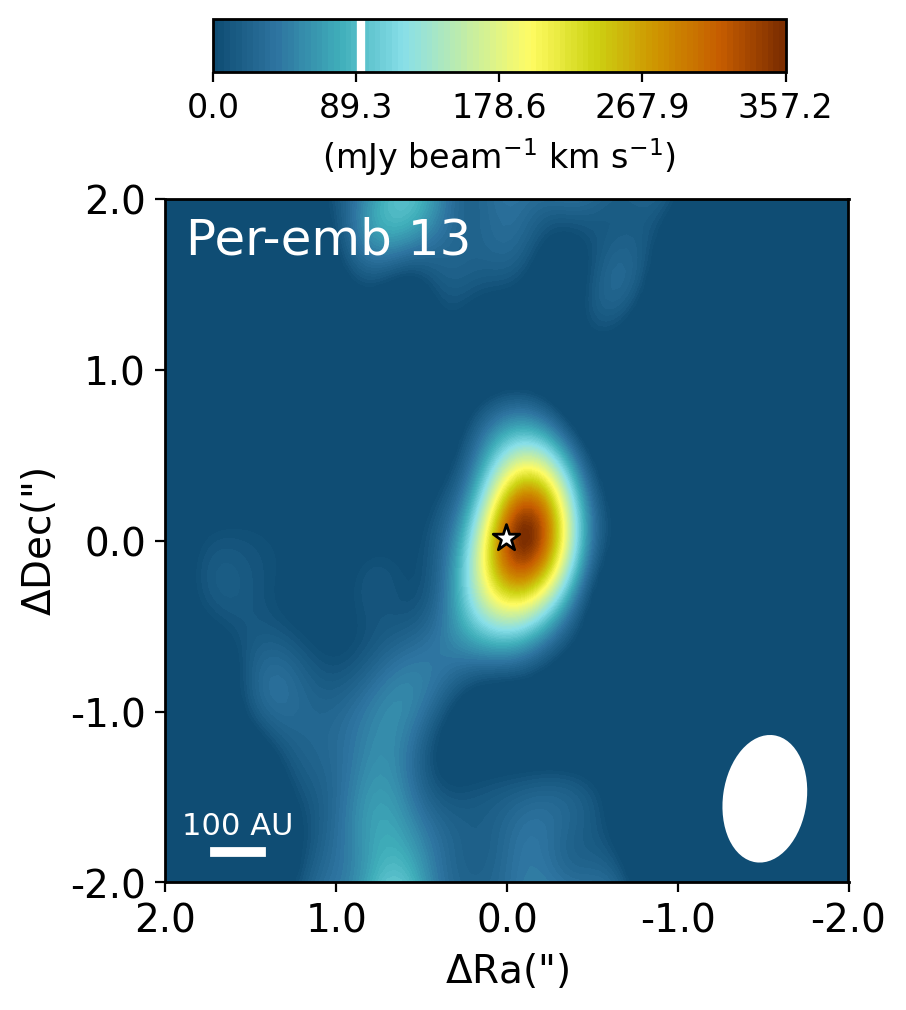}
\includegraphics[width=0.325\linewidth]{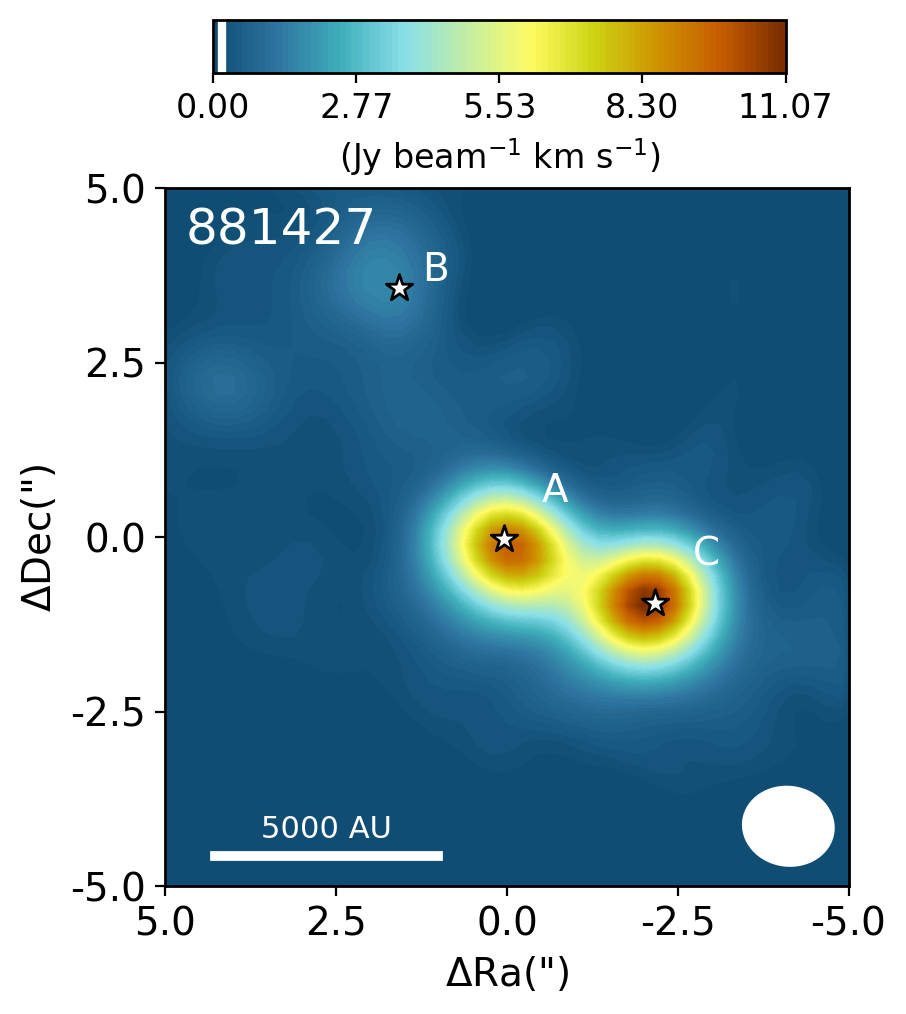}
\caption{
Integrated intensity maps of the CH$_3$OH $2_{1,1}-1_{0,1}$ line for the low-mass protostar B1-c (left), CH$_3$OH $5_{1,4}-4_{1,3}$ line for the low-mass protostar Per-emb 13 from PEACHES (middle), and the CH$_3$OH $8_{0,8}-7_{1,6}$ line for the high-mass 881427 cluster (right). The color scale is shown on top of each image. The images are integrated over [-5,5]~\kms\ with respect to the $V_\mathrm{lsr}$. The white vertical line in the colorbar indicates the $3\sigma$ threshold. The peaks of the continuum are indicated with the white stars. The white ellipse in the lower right of each image depicts the beam size and in the lower left a physical scale bar is displayed.
}
\label{fig:example_maps}
\end{figure*}

In total 50 protostars in Perseus were observed in the Perseus ALMA Chemistry Survey (PEACHES) in programs 2016.1.01501.S and 2017.1.01462.S (PI: N. Sakai). The goal of this program is to probe the complex chemistry toward embedded protostars in Perseus at $\sim0.4^{\prime\prime}$ (i.e., 150~au) scales. The full list of PEACHES sources is presented in Table~\ref{tab:observations}. The reduction and imaging of the PEACHES data are described in detail by \citet{Yang2021}. Only two transitions of CH$_{3}^{18}$OH are covered at low spectral resolution ($\sim 1.2$~\kms). Furthermore, the PEACHES data contain three transitions of $^{13}$CH$_3$OH of which the 259.03649~GHz line is blended with HDCO for most COM-rich sources. Several strong transitions of the main CH$_3$OH isotopologue are covered which likely produce optically thick lines. All PEACHES sources were also observed in the VLA Nascent Disk and Multiplicity Survey (VANDAM) showing that about $\sim50$~\% of the Class~0 and $\sim20$~\% of the Class I sources are part of multiple systems \citep{Tobin2016}. Moreover, the presence of disks at $>10$~au scales around several Perseus sources was presented by \citet{Segura-Cox2018}.

The ALMA Evolutionary study of High Mass Protocluster Formation in the Galaxy (ALMAGAL) survey (2019.1.00195.L; PI: S. Molinari) is a large program targeting over 1000 dense clumps with $M > 500$~M$_\odot$ and $d < 7.5$~kpc in Band~6. The ALMAGAL sources were targeted based on compact sources within the {\it Herschel} Hi-Gal survey \citep{Molinari2010,Elia2017,Elia2021}. In this paper, a subsample of 40 high-mass protostellar cores is selected based on a combination of high bolometric luminosity ($\gtrsim1000$~L$_\odot$) and richness in spectral lines. Moreover, only archival data with a beam size smaller than $2^{\prime\prime}$ ($\sim1000-5000$~au) that were public in February 2021 are included. As discussed below, this selection introduces a bias. The data are pipeline calibrated and imaged with the Common Astronomy Software Applications\footnote{\url{https://casa.nrao.edu/}} \citep[CASA;][]{McMullin2007} version 5.6.1. The ALMAGAL data cover multiple transitions of (likely) optically thick CH$_3$OH, four transitions of $^{13}$CH$_3$OH, and nine transitions of CH$_{3}^{18}$OH. Especially for the line rich sources with broad lines, many of these lines are unfortunately affected by line blending. 

The integrated intensity maps of CH$_3$OH are presented in Fig.~\ref{fig:example_maps} for a few representative sources: B1-c from the 2017.1.01174.S program, Per-emb 13 (NGC~1333~IRAS~4B) from PEACHES, and 881427 from ALMAGAL. For B1-c and Per-emb~13, some outflow emission is evident, but this does not influence the analysis below. The position of the continuum emission peak for all sources is listed in Table~\ref{tab:observations} and the peak continuum flux within the central beam for each source is listed in Table~\ref{tab:properties_quantities}. All spectra are extracted from the central pixel on the continuum peak positions. 

Several of the sources used in this work (e.g., 881427 in Fig.~\ref{fig:example_maps}) are resolved into small clusters in the higher-resolution ALMA images (see Table~\ref{tab:observations}). However, their luminosity is often estimated from {\it Herschel} observations within a $\sim15^{\prime\prime}$ beam \citep{Murillo2016,Elia2017,Elia2021} and therefore luminosity estimates of the individual cores are unknown. In such cases, as a zeroth-order approximation, the luminosity of each individual core is estimated by dividing the luminosity over the multiple sources. The fraction attributed to each source is computed as the peak continuum flux of the corresponding core divided by the sum of peak continuum fluxes from all cores.

\subsection{Deriving the column density}
The column density of CH$_3$OH, $N_\mathrm{CH_3OH}$, is calculated from the spectrum through local thermodynamic equilibrium (LTE) models using the spectral analysis tool, CASSIS\footnote{\url{http://cassis.irap.omp.eu/}} \citep{Vastel2015}. Since spectral lines originating from the main isotopologue of CH$_3$OH are likely optically thick, lines from optically thin isotopologues such as $^{13}$CH$_3$OH and CH$_{3}^{18}$OH need to be invoked to get an accurate estimate of the column density. The $\mathrm{^{12}C/^{13}C}$ and $\mathrm{^{16}O/^{18}O}$ ratios are dependent on the Galactocentric distance and are determined using the relations of \citet{Milam2005} and \citet{Wilson1994}, respectively. For the local interstellar medium (ISM), this results in ratios of $\mathrm{^{12}C/^{13}C} \sim 70$ and $\mathrm{^{16}O/^{18}O} \sim 560$. The full line lists are acquired from the CDMS catalog\footnote{\url{https://cdms.astro.uni-koeln.de/}} \citep{Muller2001,Muller2005,Endres2016} and are presented for each dataset in Appendix~\ref{app:CH3OH_transitions}. 

For the PEACHES and ALMAGAL datasets, generally only a few lines of each isotopologue are detected. Therefore, for simplicity, an excitation temperature of $T_\mathrm{ex}=150$~K is assumed, which is roughly the average of what is observed toward both low-mass and high-mass protostellar systems \citep[typical values lie in the range of $100-300$~K, e.g.,][]{Bogelund2018,Bogelund2019,vanGelder2020,Yang2021}. Changing the excitation temperature within the $100-300$~K range leads to only a factor $\sim2$ variation in the derived column densities since lines with a range of $E_\mathrm{up}$ from $\sim50$~K to $\sim800$~K are available for the analysis. The column density $N$ of each isotopologue is derived separately using a similar grid fitting method as \citet{vanGelder2020} with $N$ and the full width at half maximum (FWHM) of the line as free parameters. The size of the emitting region is set equal to the beam size and is presented for each source in Table~\ref{tab:properties_quantities}. To exclude emission associated with outflows, only narrow lines (FWHM$\lesssim$~few~km~s$^{-1}$) with $E_\mathrm{up}\gtrsim50$~K are used in the analysis. Moreover, blended lines are excluded from the fit. The 2$\sigma$ uncertainty on $N$ is derived from the grid. Careful inspection by eye was conducted to test the validity of the fits and derived column densities.

\begin{figure*}[]
\includegraphics[height=4.9cm]{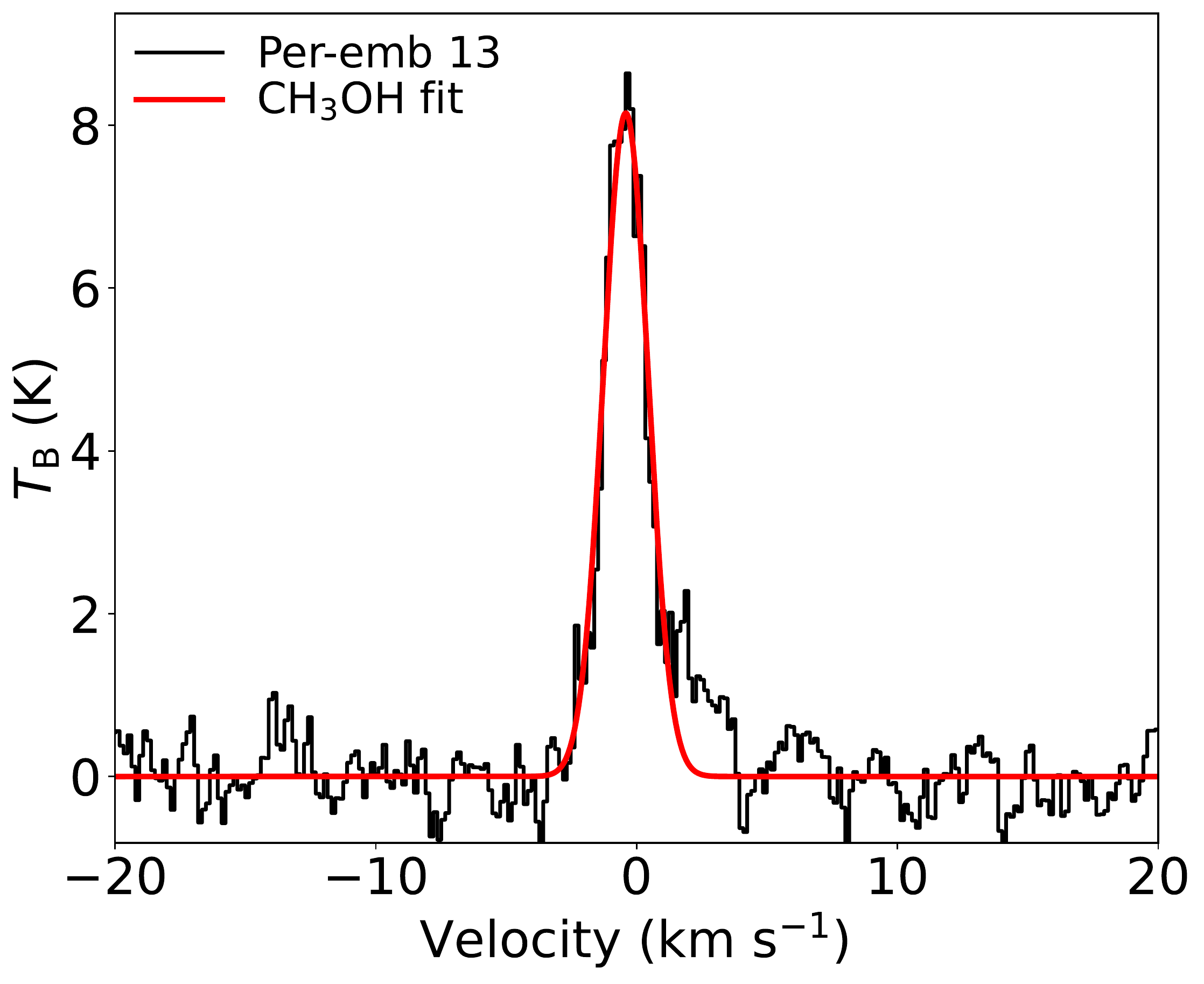}
\includegraphics[height=4.9cm]{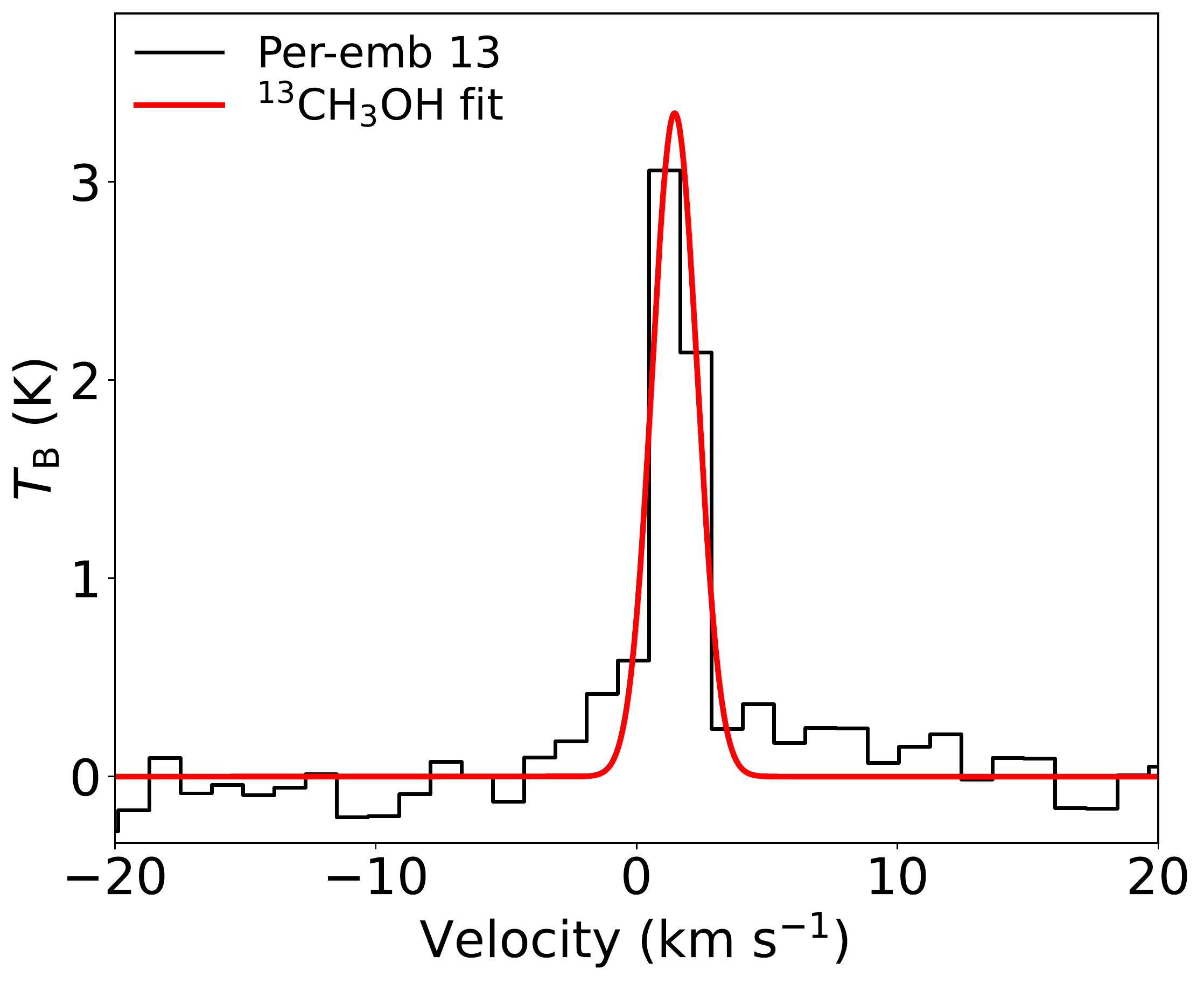}
\includegraphics[height=4.9cm]{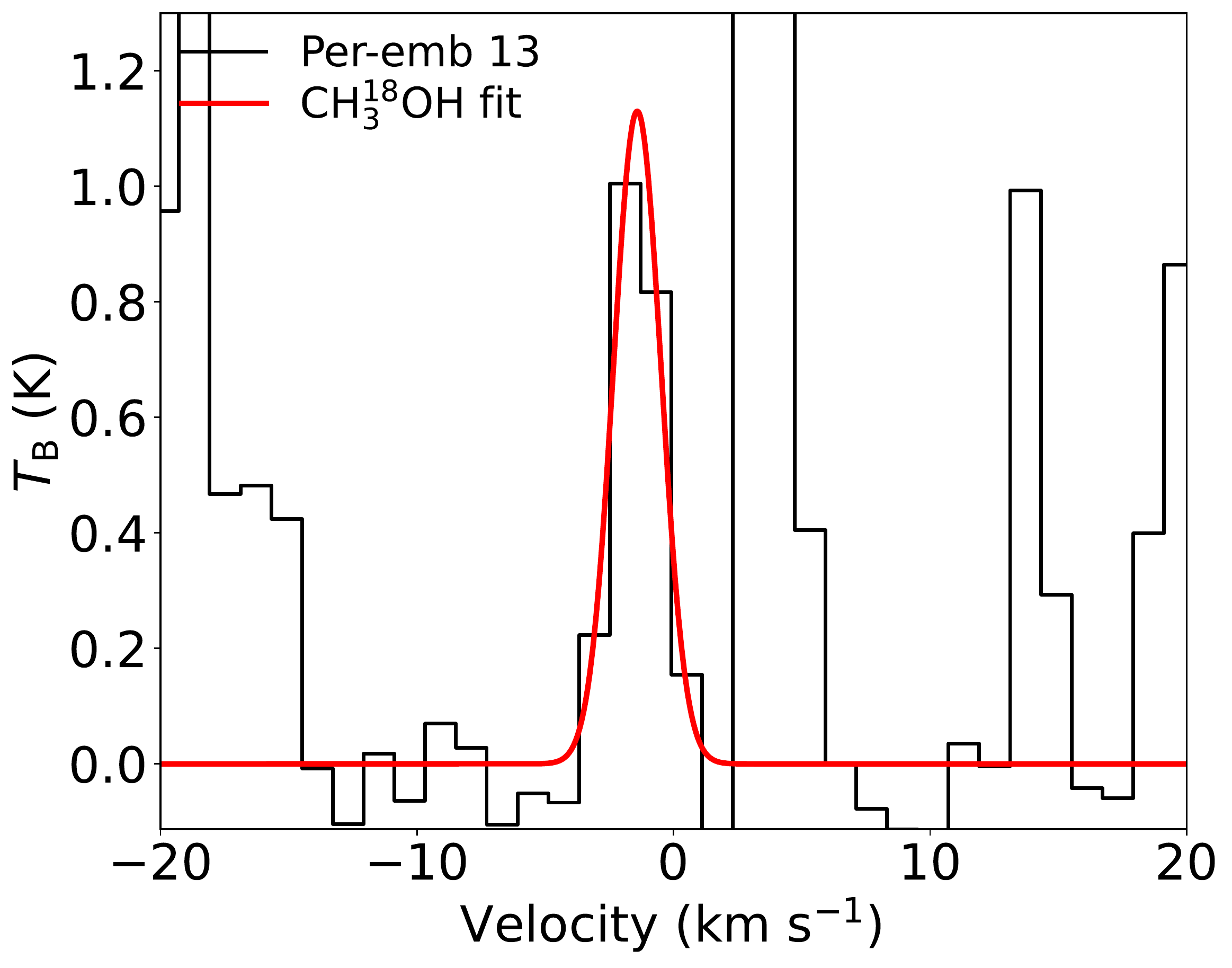}
\includegraphics[height=4.9cm]{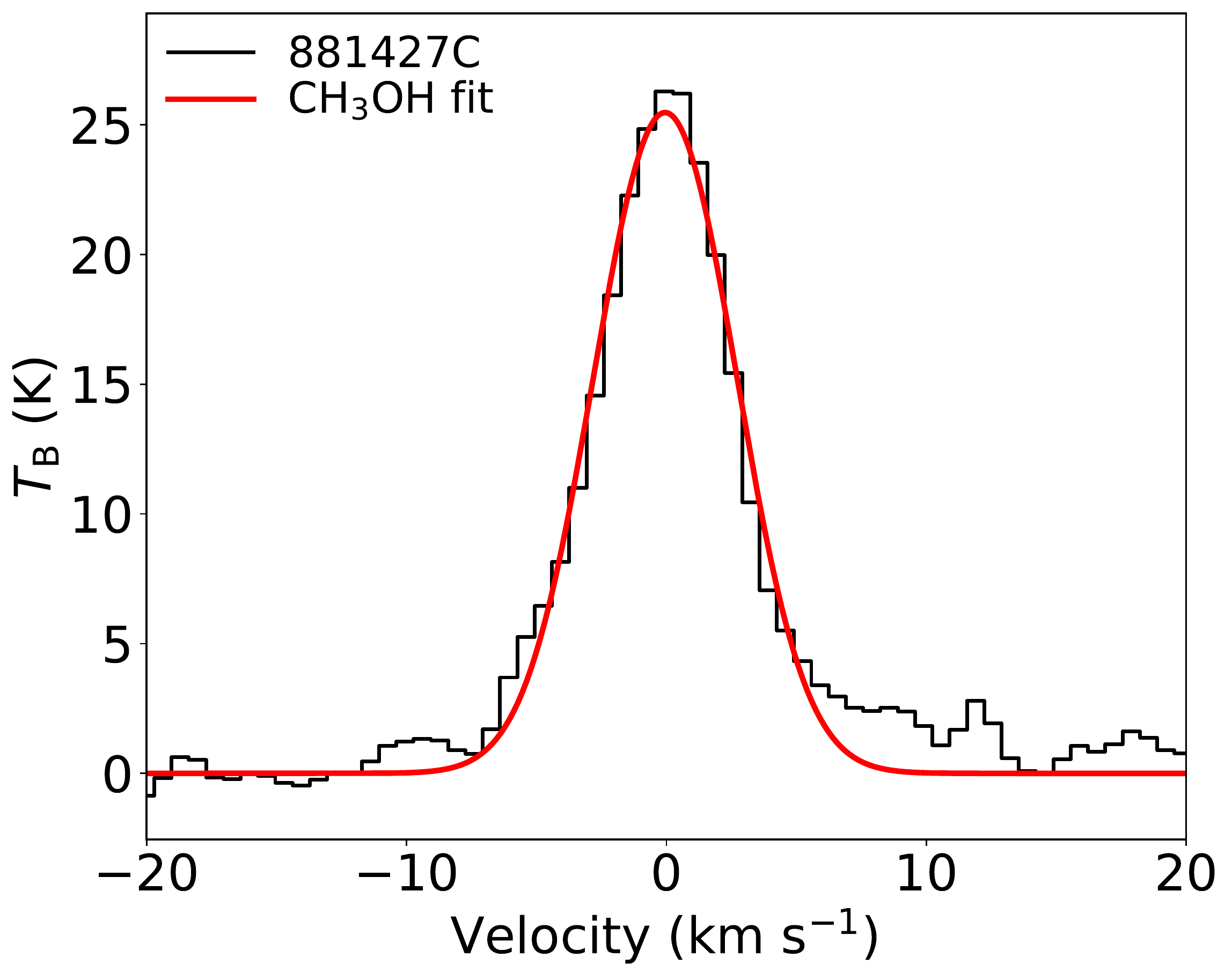}
\includegraphics[height=4.9cm]{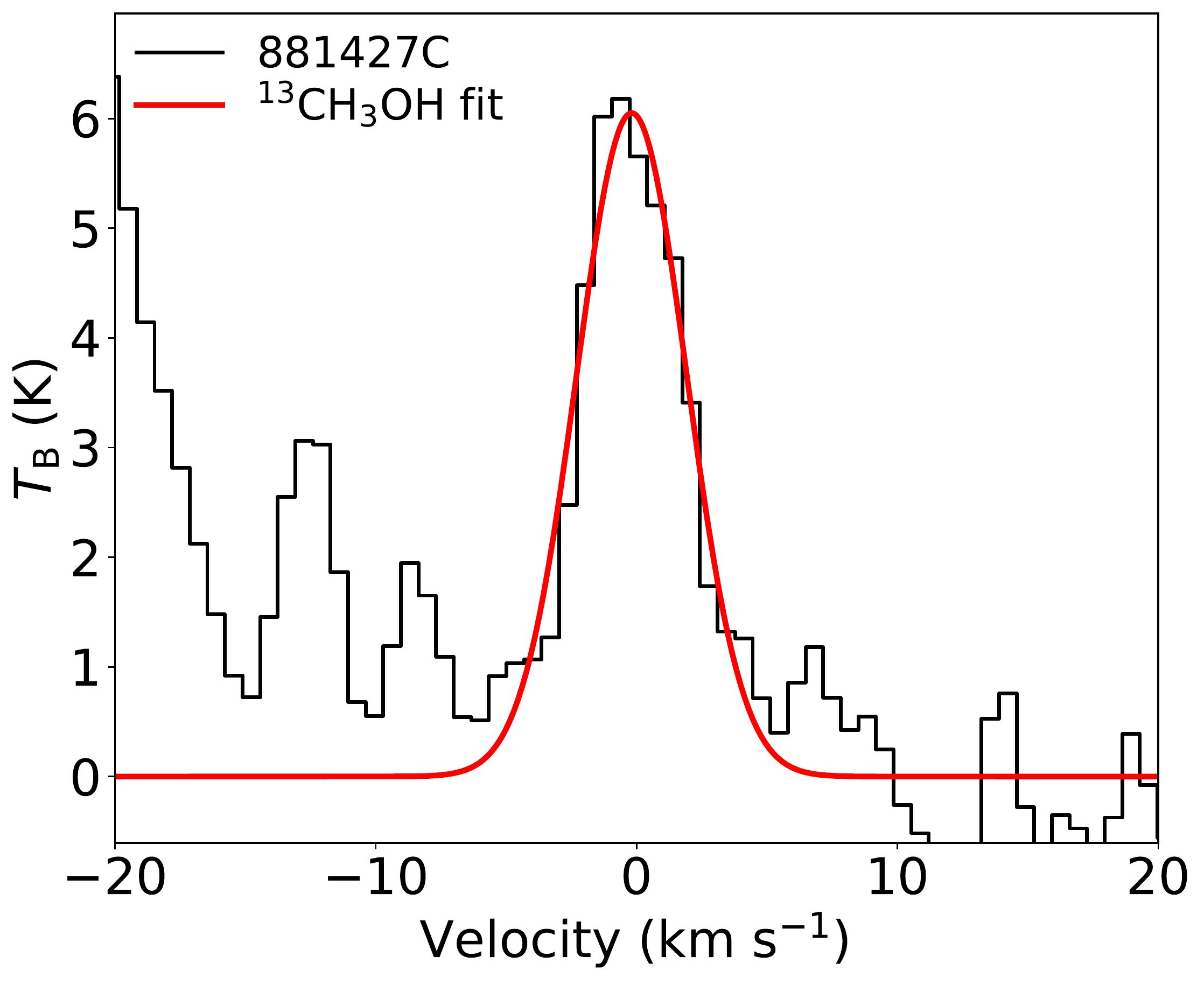}
\includegraphics[height=4.9cm]{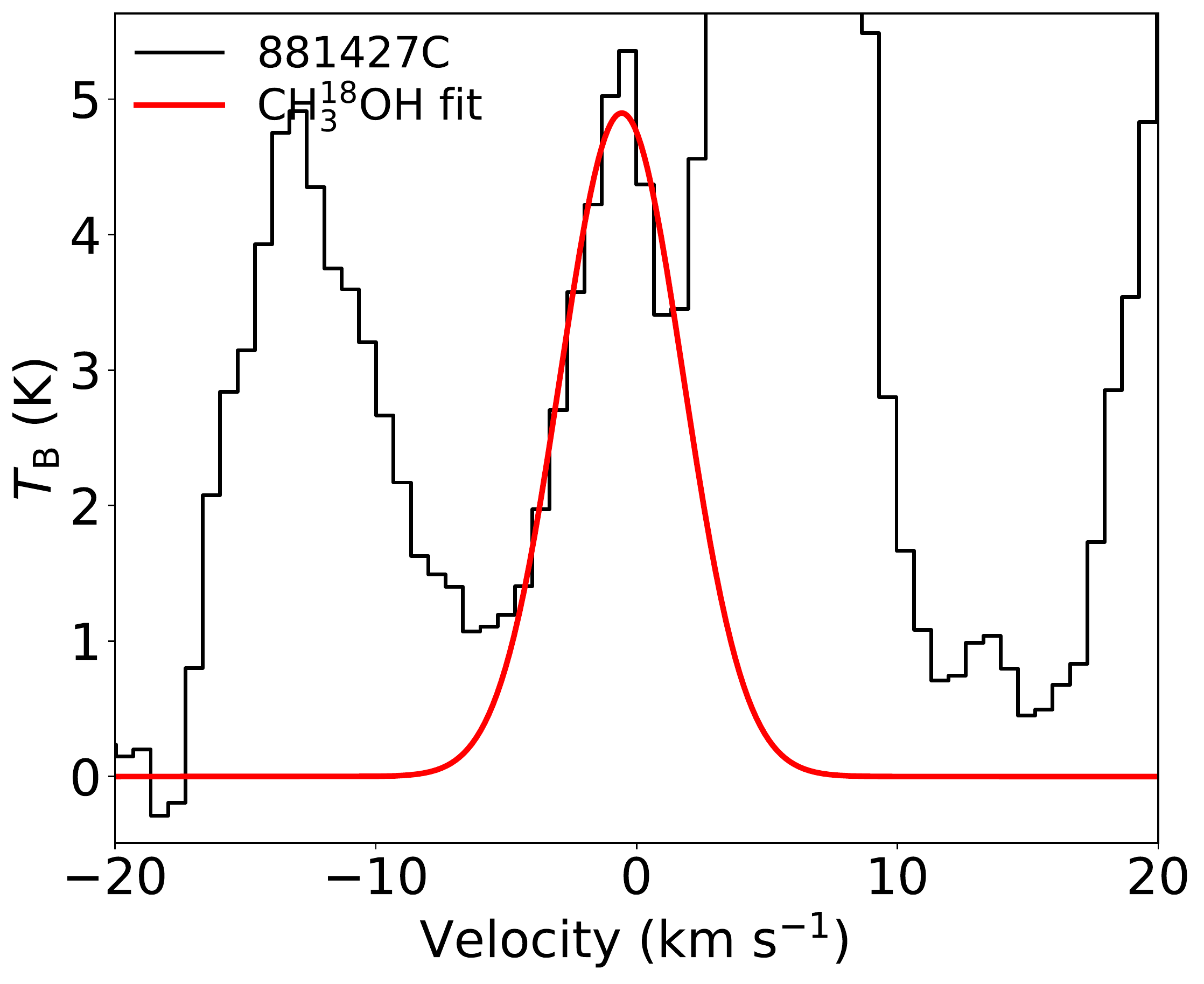}
\caption{
Spectral line fits of CH$_3$OH (left), $^{13}$CH$_3$OH (middle), and CH$_3^{18}$OH (right) for Per-emb~13 from the PEACHES sample (top row) and 881427C from the ALMAGAL sample (bottom row). The data corrected for the $V_\mathrm{lsr}$ are shown in black and the fit for $T_\mathrm{ex}=150$~K is shown in red. All lines are $>3\sigma$ detections. The top row shows the CH$_3$OH $5_{1, 4}-4_{1, 3}$ ($E_\mathrm{up}=50$~K), $^{13}$CH$_3$OH $23_{3,20}-23_{2,21}$ ($E_\mathrm{up}=675$~K), and CH$_3^{18}$OH $11_{2,10}-10_{3,7}$ ($E_\mathrm{up}=184$~K) lines, respectively, and the bottom row shows the CH$_3$OH $8_{0,8}-7_{1,6}$ ($E_\mathrm{up}=97$~K), $^{13}$CH$_3$OH $14_{1,13}-13_{2,12}$ ($E_\mathrm{up}=254$~K), and 
CH$_3^{18}$OH $8_{1,8}-7_{0,7}$ ($E_\mathrm{up}=86$~K) lines.
}
\label{fig:example_fits}
\end{figure*}

In Fig.~\ref{fig:example_fits}, fits to lines of CH$_3$OH and its isotopologues are presented for Per-emb~13 from the PEACHES sample and 881427C from the ALMAGAL sample. All lines in Fig.~\ref{fig:example_fits} are $>$3$\sigma$ detections. The CH$_3^{18}$OH lines suffer from line blending in 881427C as well as many other PEACHES and ALMAGAL sources. Since all other lines originating from $^{13}$CH$_3$OH in the PEACHES data suffer from line blending (e.g., with HDCO), the $^{13}$CH$_3$OH $23_{3,20}-23_{2,21}$ ($E_\mathrm{up}=675$~K) line in Fig.~\ref{fig:example_fits} often provides the only constraint on the column density. 

The derived CH$_3$OH column densities are presented in Table~\ref{tab:properties_quantities}. The column densities of the sources in the 2017.1.01174.S and 2017.1.01350.S datasets were derived by \citet{vanGelder2020}. Only the Band~6 results are used in this work. When no lines originating from CH$_3^{18}$OH are detected, the column density of CH$_3$OH is derived from $^{13}$CH$_3$OH lines. In case that only upper limits on the column densities of both $^{13}$CH$_3$OH and CH$_3^{18}$OH could be derived, $N_\mathrm{CH_3OH}$ is calculated by setting the 3$\sigma$ upper limit based on scaling the 3$\sigma$ upper limit of $^{13}$CH$_3$OH and the lower limit based on the main isotopologue. This situation results in rather large errorbars on $N_\mathrm{CH_3OH}$ for 9 low-mass and 24 high-mass sources. When the main isotopologue of CH$_3$OH is also not detected, the $3\sigma$ upper limit is reported. The upper limits on $N_\mathrm{CH_3OH}$ are not homogeneous across the full sample since they are derived from various ALMA programs with a range of sensitivities covering different transitions of CH$_3$OH with different upper energy levels and Einstein $A_\mathrm{ij}$ coefficients. For sources with detections, this inhomogeneity is not present except for the uncertainty based on the assumed $T_\mathrm{ex}$. 

Besides the sources presented in Table~\ref{tab:observations}, also other sources with or without warm (\mbox{$T\gtrsim100$~K}) CH$_3$OH detections are included from the literature. These sources include well-known low-mass hot cores such as IRAS~16293-2422 \citep{Jorgensen2016,Jorgensen2018,Manigand2020}, NGC~1333~IRAS4A \citep{deSimone2020}, BHR~71 \citep{Yang2020}, and L438 \citep{Jacobsen2019}, but also sources with disks showing emission of COMs such as HH~212 \citep{Lee2017_HH212,Lee2019_HH212} and outbursting sources such as SVS~13A \citep{Bianchi2017,Hsieh2019} and V883~Ori \citep{vantHoff2018,Lee_V883_2019}. Also non-detections of CH$_3$OH such as found for several Class~I sources in Ophiuchus are included \citep{ArturdelaVillarmois2019}. Moreover, classical high-mass hot cores such as AFGL~4176 \citep{Bogelund2019}, NGC~6334I \citep{Bogelund2018}, and Sgr~B2 \citep{Belloche2013,Muller2016,Bonfand2017} are taken into account. A 20\% uncertainty on $N_\mathrm{CH_3OH}$ is assumed for literature sources where no uncertainty on the column density was reported. The full list of literature sources is also presented in Table~\ref{tab:properties_quantities}. Including the 36 literature sources, the total sample studied in this work contains 184 unique sources (some sources such as B1-bS and Per-emb~44 (SVS~13A) are covered in both our data and literature studies).

\begin{SCfigure*}[][]
\includegraphics[width=1.5\linewidth]{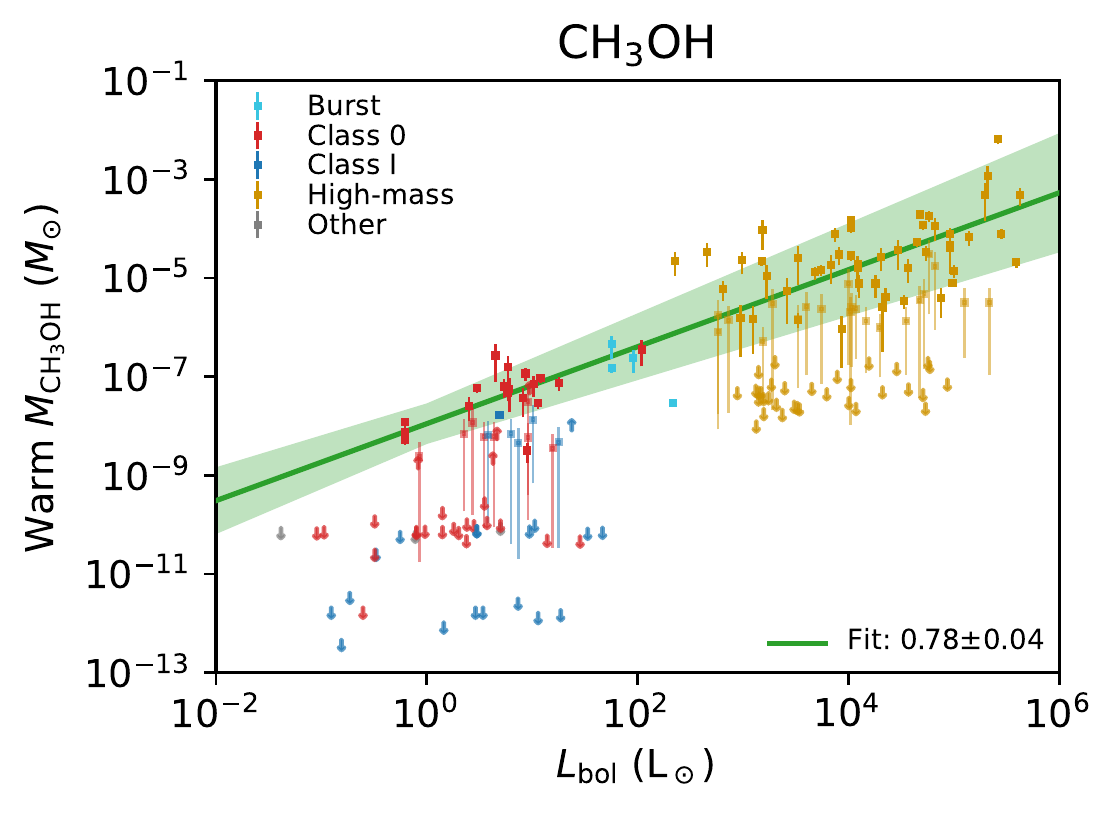}
\caption{
Warm CH$_3$OH mass as function of the bolometric luminosity. Different colors denote different observational classes, where all sources with $L_\mathrm{bol}>1000$~L$_\odot$ are classified as high-mass sources. Additionally, sources which are suggested to currently be in a burst phase \citep[e.g., V883~Ori and IRAS~2A;][]{vantHoff2018,Lee_V883_2019,Hsieh2019} are highlighted. Sources in the "other" category include Class II and flat spectrum sources. The green line indicates the best-fit power-law model to the data points (excluding the upper limits) and the green shaded area the 3$\sigma$ uncertainty on the fit. 
}
\label{fig:M_vs_L_Classes}
\end{SCfigure*}

\subsection{Calculating the warm methanol mass}
\label{subsec:calc_warm_methanol_mass}
In order to get a measurement of the amount of warm (\mbox{$T\gtrsim100$~K}) CH$_3$OH gas, observational dependencies should be taken into account. Column densities provide a measure of the amount of CH$_3$OH, but these depend on the assumed size of the emitting region. From the 184 sources studied in this work, $\sim$30 (mostly high-mass) sources show spatially resolved CH$_3$OH emission, but nevertheless in these cases most of the CH$_3$OH emission is still located within the central beam. Furthermore, emission from $^{13}$CH$_3$OH and CH$_3^{18}$OH, that gives more stringent constraints on $N_\mathrm{CH_3OH}$, is only spatially resolved in a few high-mass sources. Therefore, in this work the size of the emitting region is assumed to be equal to the size of the beam. Because our sample consists of many different types of sources at a range of distances taken with different ALMA programs, all derived column densities are multiplied with the physical area of the beam to derive the total number of CH$_3$OH molecules $\mathcal{N}_\mathrm{CH_3OH}$ within the beam,
\begin{align}
\mathcal{N}_\mathrm{CH_3OH} = N_\mathrm{CH_3OH} \pi R_\mathrm{beam}^2.
\end{align}
Here, $N_\mathrm{CH_3OH}$ is the observed column density in the beam and $R_\mathrm{beam}$ is the physical radius of the beam,
\begin{align}
R_\mathrm{beam} = d \tan\left(\frac{\theta_\mathrm{beam}}{2}\right),
\label{eq:R_source}
\end{align}
with $d$ the distance to the source and $\theta_\mathrm{beam}$ the angular size of the beam. For some of the literature sources, the adopted emitting region is different from the beam size \citep[e.g.,][]{Bianchi2017,Jacobsen2019}. In these cases, their assumed size of the emitting region is adopted in the computation of $\mathcal{N}_\mathrm{CH_3OH}$ (see $\theta_\mathrm{source}$ column in Table~\ref{tab:properties_quantities}). However, it is important to note that the assumed size of the emitting region does not alter the resulting value of $\mathcal{N}_\mathrm{CH_3OH}$ as long as the beam averaged column density is derived from optically thin lines (i.e., from the $^{13}$C or $^{18}$O isotopologues). If the lines are optically thick, this approach provides a lower limit on $\mathcal{N}_\mathrm{CH_3OH}$. Finally, the warm gaseous CH$_3$OH mass $M_\mathrm{CH_3OH}$ is computed through,
\begin{align}
M_\mathrm{CH_3OH} = \mathcal{N}_\mathrm{CH_3OH} m_\mathrm{CH_3OH},
\end{align}
where  $m_\mathrm{CH_3OH}$ is the molar mass of a methanol molecule of $5.32 \times 10^{-23}$~gr or $2.67 \times 10^{-56}$~M$_\odot$.

\section{Results}
\label{sec:results}
\subsection{Amount of warm methanol from low to high mass}
\label{subsec:warm_methanol_mass}
The derived values of $M_\mathrm{CH_3OH}$ are presented in Table~\ref{tab:properties_quantities} for all sources. In Fig.~\ref{fig:M_vs_L_Classes}, $M_\mathrm{CH_3OH}$ is plotted as function of the bolometric luminosity $L_\mathrm{bol}$. A clear trend of $M_\mathrm{CH_3OH}$ as function of $L_\mathrm{bol}$ is evident where more luminous sources have more CH$_3$OH in the gas phase. Excluding upper limits, a simple power-law fit to the results gives a $M_\mathrm{CH_3OH} \propto L_\mathrm{bol}^{0.78\pm0.04}$ relation which is presented with green line in Fig.~\ref{fig:M_vs_L_Classes}. The 3$\sigma$ uncertainty on the fit is shown as the green shaded area. The positive correlation found here agrees well with that found for low-mass Class~0 protostars in Orion \citep{Hsu2022}. The positive correlation is in line with the expectation that $M_\mathrm{CH_3OH}$ increases with $L_\mathrm{bol}$ since sources with a higher luminosity are expected to have their CH$_3$OH snowline at larger radii which results in more CH$_3$OH in the gas phase. Furthermore, the envelope of sources with higher $L_\mathrm{bol}$ are more massive and thus contain more CH$_3$OH mass in the first place. However, the result of the fit strongly depends on the difference in warm $M_\mathrm{CH_3OH}$ between the low-mass ($L_\mathrm{bol} \lesssim 100$~L$_\odot$) and high-mass ($L_\mathrm{bol} \gtrsim 1000$~L$_\odot$) protostars. Fitting these two subsamples individually gives a less strong correlation between $M_\mathrm{CH_3OH}$ and $L_\mathrm{bol}$ ($M_\mathrm{CH_3OH} \propto L_\mathrm{bol}^{0.42\pm0.12}$ and $M_\mathrm{CH_3OH} \propto L_\mathrm{bol}^{0.34\pm0.13}$ for low-mass and high-mass protostars, respectively). Due to the large uncertainties on the fits of both subsamples, no significant conclusions can be made.

For the lower mass sources ($L_\mathrm{bol} \lesssim 100$~L$_\odot$) a scatter of more than four orders of magnitude is present. On average Class~0 sources seem to show more CH$_3$OH mass ($\sim10^{-7}$~M$_\odot$) than more evolved Class~I sources ($\lesssim10^{-10}$~M$_\odot$). In fact, for the majority of the Class I sources only upper limits can be derived on $M_\mathrm{CH_3OH}$ \citep[e.g., the Class I sources in Ophiuchus;][]{ArturdelaVillarmois2019}, but some Class I sources do show emission of COMs including CH$_3$OH \citep[e.g., L1551 IRS5;][]{Bianchi2020}. Moreover, known bursting sources such as IRAS~2A \citep[or Per-emb 27;][]{Hsieh2019,Yang2021} and V883~Ori \citep{vantHoff2018,Lee_V883_2019} show a large amount of CH$_3$OH mass despite the presence of disk-like structures \citep{Segura-Cox2018,Maury2019}. 

\begin{SCfigure*}[][]
\includegraphics[width=1.5\linewidth]{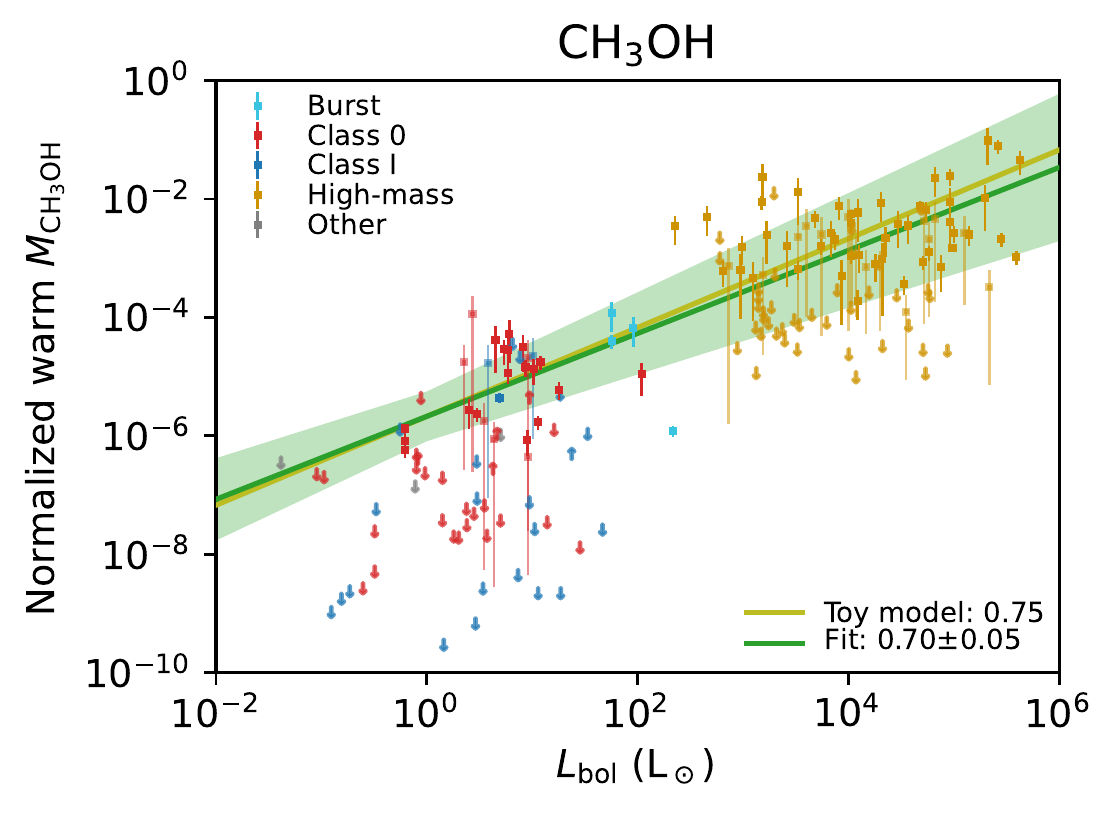}
\caption{
Normalized warm gaseous CH$_3$OH mass, $M_\mathrm{CH_3OH}/M_\mathrm{dust,0}$, as function of the bolometric luminosity. Different colors denote different observational classes, where all sources with $L_\mathrm{bol}>1000$~L$_\odot$ are classified as high-mass sources. Additionally, sources which are suggested to currently be in a burst phase \citep[e.g., V883 Ori and IRAS2A;][]{vantHoff2018,Lee_V883_2019,Hsieh2019} are highlighted. Sources in the "other" category include Class II and flat spectrum sources. The green line indicates the best-fit power-law model to the data points (excluding the upper limits) and the green shaded area the 3$\sigma$ uncertainty on the fit. The yellow line indicates the relation for the toy model of Eq.~\eqref{eq:M_model_envelope} scaled to match the best-fit power-law model. 
}
\label{fig:M_div_M_dust_vs_L_Classes}
\end{SCfigure*}

The high-mass ($L_\mathrm{bol} \gtrsim 1000$~L$_\odot$) sources show higher gaseous CH$_3$OH masses (i.e., $10^{-7}-10^{-3}$~M$_\odot$) compared to the lower-mass sources. A similar scatter of about four orders of magnitude is present, but fewer sources seem to show lower CH$_3$OH masses. This is likely a sample bias effect since mostly very line-rich sources from the ALMAGAL dataset are analyzed whereas many ALMAGAL sources show upper limits. Hence, in reality more sources with $M_\mathrm{CH_3OH} \lesssim 10^{-7}$~M$_\odot$ will be present among the high-mass sources as well. Alternatively, the emitting region in high-mass sources could be larger than typical disk sizes reached resulting in fewer sources for which the disk has a significant effect on the thermal structure.

\subsection{Comparison to spherically symmetric infalling envelope}
\label{subsec:toy_model}
To quantify the effect of the luminosity on the amount of warm methanol, a simple toy model of a spherically symmetric infalling envelope is constructed, see Appendix~\ref{app:toy_model}. In this simple toy model, the warm methanol mass is linked to $L_\mathrm{bol}$, 
\begin{align}
M_\mathrm{CH_3OH} \propto M_\mathrm{0} L_\mathrm{bol}^{3/4},
\label{eq:M_model_envelope}
\end{align}
where $M_\mathrm{0}$ is the total warm + cold mass contained within a reference radius $R_\mathrm{0}$. As qualitatively discussed in Sect.~\ref{subsec:warm_methanol_mass}, the warm methanol mass is thus dependent on both the envelope mass within a reference radius (or alternatively the density at a reference radius) and the luminosity of the source. The former is straightforward since a higher mass will result in a higher methanol mass to begin with. The latter is the result of the snowline of CH$_3$OH moving to larger radii for larger luminosities. 

Hence, to probe the effect of the luminosity on the snowline radius in our observations, the values of $M_\mathrm{CH_3OH}$ presented in Sect.~\ref{subsec:warm_methanol_mass} should be divided by a reference mass $M_\mathrm{0}$ defined within a reference radius $R_\mathrm{0}$. The protostellar envelope mass $M_\mathrm{env}$ is not a good option as $M_\mathrm{0}$ since it is not consistently constrained for many sources and depends on the adopted outer radius \citep[e.g.,][]{Kristensen2012}. Therefore, in this work the dust mass within a common arbitrary radius $R_0 = 200$~au, denoted as $M_\mathrm{dust,0}$, is used as the reference mass. The computation of $M_\mathrm{dust,0}$ from the continuum flux is detailed in Appendix~\ref{app:M_dust,0} and the derived values are reported in Table~\ref{tab:properties_quantities}. The choice for $R_0 = 200$~au is based on it being larger than the typical disk size of low-mass protostars. Despite 200~au being smaller than the typical size of a high-mass hot core, we do not expect this to influence our analysis since the scale on which the continuum flux is measured ($>1000$~au) is dominated by the envelope. Furthermore, the continuum flux on 200~au scales is not filtered out in both the PEACHES and ALMAGAL observations.

In Fig.~\ref{fig:M_div_M_dust_vs_L_Classes}, the normalized warm gaseous CH$_3$OH mass $M_\mathrm{CH_3OH}/M_\mathrm{dust,0}$ is presented as function of $L_\mathrm{bol}$. It is important to note that the $M_\mathrm{CH_3OH}/M_\mathrm{dust,0}$ ratio is a dimensionless quantity and does not represent an abundance of CH$_3$OH in the system. The warm gaseous CH$_3$OH mass is derived from a region taken to be similar in size as the beam. The dust mass $M_\mathrm{dust,0}$ includes both cold and warm material in the disk and envelope and is defined as the mass within the fixed radius $R_0 = 200$~au. For most low-mass sources, the snowline is expected to be well within 200~au whereas for high-mass sources the snowline is at radii larger than 200~au. However, changing the normalization to larger or smaller $R_0$ does not affect the results discussed below.

Despite $M_\mathrm{CH_3OH}/M_\mathrm{dust,0}$ not representing an abundance of CH$_3$OH, it can be seen as a lower limit on the abundance of CH$_3$OH in the hot cores of sources where the expected snowline radius lies inward of 200~au (i.e., when $L_\mathrm{bol} \lesssim100$~L$_\odot$). For typical values of $M_\mathrm{CH_3OH}/M_\mathrm{dust,0} \sim10^{-5}$, this implies abundances of $\gtrsim10^{-7}$ with respect to H$_2$ (assuming a gas-to-dust mass ratio of 100) which is in agreement with the CH$_3$OH ice abundances in protostellar envelopes \citep[$\sim10^{-6}$ with respect to H$_2$;][]{Boogert2008,Bottinelli2010,Oberg2011}.

\begin{SCfigure*}[][]
\includegraphics[width=1.5\linewidth]{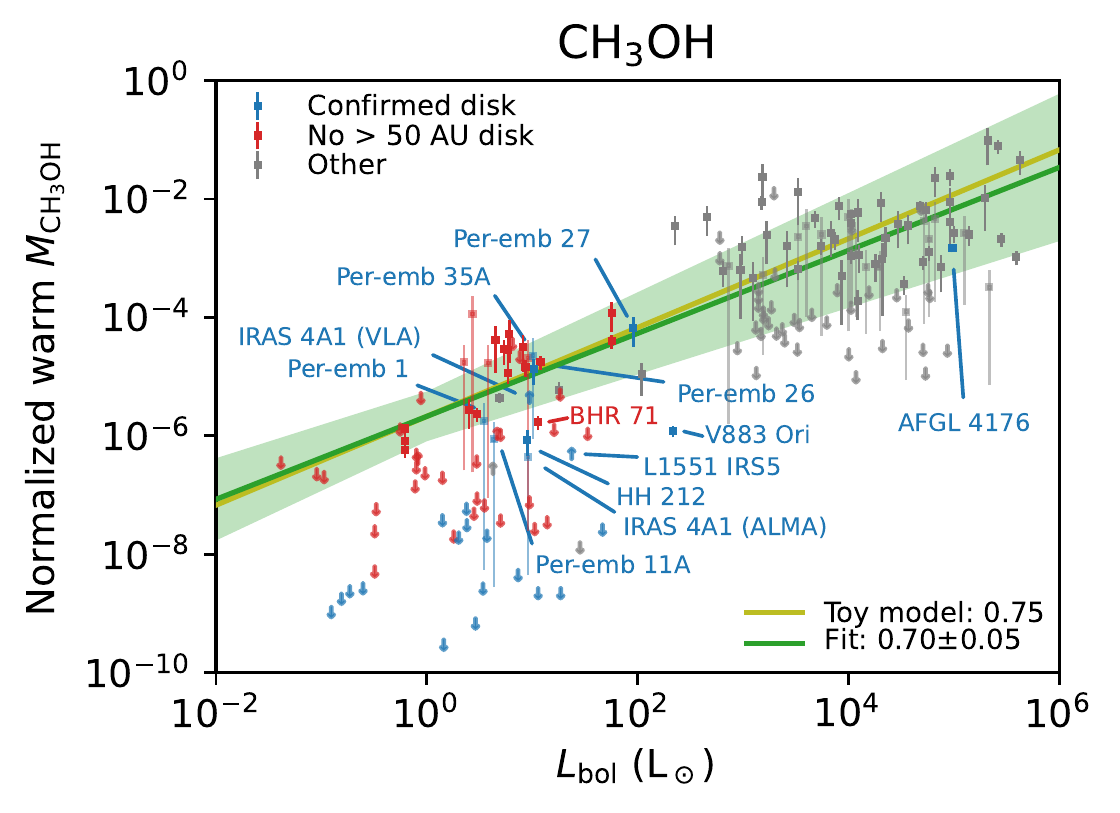}
\caption{
Normalized warm gaseous CH$_3$OH mass with respect to $M_\mathrm{dust}$ within a 200~au radius as function of the bolometric luminosity. Sources around which a disk has been detected on $>10$~au scales are indicated in blue and sources where no disk is confirmed on $>50$~au scales are indicated in red. All other sources (i.e., with no information about a disk) are shown in gray. The green line indicates the best-fit power-law model to the data points (excluding the upper limits) and the green shaded area the 3$\sigma$ uncertainty on the fit. The yellow line indicates the relation for the toy model of Eq.~\eqref{eq:M_model_envelope} scaled to match the best-fit power-law model. 
}
\label{fig:M_div_M_dust_vs_L_Diskdetections}
\end{SCfigure*}

A positive correlation of $M_\mathrm{CH_3OH}/M_\mathrm{dust,0}$ with $L_\mathrm{bol}$ is evident, $M_\mathrm{CH_3OH}/M_\mathrm{dust,0} \propto L_\mathrm{bol}^{0.70\pm0.05}$ (again excluding upper limits). The slope of the power law is in good agreement with the slope derived in Eq.~\eqref{eq:M_model_envelope} for the simple toy model (yellow line in Fig.~\ref{fig:M_div_M_dust_vs_L_Classes} scaled to match the best-fit power-law mode). This indicates that for sources with high $M_\mathrm{CH_3OH}/M_\mathrm{dust,0}$ the thermal structure of the envelope is likely not affected by the disk. Since the sources are corrected for the source's mass, the increase in the $M_\mathrm{CH_3OH}/M_\mathrm{dust,0}$ ratio originates from the snowline moving further out in systems with higher luminosity. In turn, this leads to more gaseous warm CH$_3$OH mass. Therefore, the agreement with the toy model can be considered as an observational confirmation of the scaling between the sublimation radius (i.e., snowline radius) and source luminosity of $R_\mathrm{sub} \propto L_\mathrm{bol}^{1/2}$ found in radiative transfer calculations \citep[i.e., Eq.~\eqref{eq:snowline_Lbol_relation},][]{Bisschop2007,vantHoff2022}.

The correlation between $M_\mathrm{CH_3OH}/M_\mathrm{dust,0}$ and $ L_\mathrm{bol}$ is mostly evident if both the low-mass and high-mass samples are combined. Among these two subsamples individually a positive correlation is less evident: $M_\mathrm{CH_3OH}/M_\mathrm{dust,0} \propto L_\mathrm{bol}^{0.38\pm0.17}$ and $M_\mathrm{CH_3OH}/M_\mathrm{dust,0} \propto L_\mathrm{bol}^{0.15\pm0.10}$ for the low-mass and high-mass protostars, respectively. However, the uncertainties on the fits are large due to the large scatter within each subsample. Moreover, the fits are more sensitive to individual datapoints (e.g., B1-bS at low $L_\mathrm{bol}$). Therefore, no significant conclusions can be made.

Similarly to Fig.~\ref{fig:M_vs_L_Classes}, a scatter of more than four orders of magnitude is evident in Fig.~\ref{fig:M_div_M_dust_vs_L_Classes} among both the low-mass and high-mass sources. Likewise, on average the Class I sources show lower values  ($M_\mathrm{CH_3OH}/M_\mathrm{dust,0} \lesssim10^{-6}$) than the (younger) Class~0 sources ($M_\mathrm{CH_3OH}/M_\mathrm{dust,0} \sim10^{-5}$). The large scatter indicates that the normalized gaseous CH$_3$OH mass cannot be solely explained by the toy model of a spherically symmetric infalling envelope. 

\section{Discussion}
\label{sec:discussion}
\subsection{Importance of source structure}
If all embedded protostellar systems would behave solely as spherically symmetric infalling envelopes, a larger luminosity for a given mass of the source would result in moving the CH$_3$OH snowline further out and thus in more CH$_3$OH in the gas phase. However, the large scatter in Fig.~\ref{fig:M_div_M_dust_vs_L_Classes} indicates that other mechanisms affect the gaseous CH$_3$OH mass. One possible explanation could be the presence of a disk that alters the temperature structure of the system and thus lowers the amount of warm CH$_3$OH. Alternatively, continuum optical depth can hide the emission of COMs. These cases could explain the deviation from the toy model relation of Eq.~\eqref{eq:M_model_envelope}. 

In Fig.~\ref{fig:M_div_M_dust_vs_L_Diskdetections}, the normalized warm gaseous CH$_3$OH mass is presented as function of $L_\mathrm{bol}$ similarly to Fig.~\ref{fig:M_div_M_dust_vs_L_Classes}, but now with colors indicating whether the source has a confirmed disk on $>10$~au scales (blue) or whether the absence of a $>50$~au disk is confirmed. All sources where no information on the presence of a disk is available (e.g., high-mass sources) are set in the "other" category. Besides the sources with well-known disks such as HH~212 \citep[e.g.,][]{Lee2017_HH212,Lee2019_HH212}, the presence of a disk is indicated by Keplerian rotation \citep[e.g., L1527 and V883~Ori;][]{Tobin2012,Lee_V883_2019} or based on elongated dust continuum \citep[e.g.,][]{Segura-Cox2018,Maury2019}. 

Among the low-mass sources, the presence of a disk does not directly imply that no significant COM emission can be present, given that the emission of COMs can be in the part of the envelope not shadowed by the disk or in the warm upper atmosphere of the disk \citep[e.g., HH~212;][]{Lee2017_HH212,Lee2019_HH212}. Excluding the lower limits for IRAS~4A1 \citep[Per-emb~12A;][]{deSimone2020} and L1551~IRS5 \citep{Bianchi2020}, the only seven sources in our sample with both a disk confirmed and a CH$_3$OH detection are HH~212, Per-emb~27 (IRAS~2A), V883~Ori, Per-emb~35A (IRAS~1A), Per-emb~26 (L1448-mm), Per-emb~1 (HH211), and Per-emb~11A (IC~348~MMS). Their normalized warm CH$_3$OH masses as well as their estimated dust disk radii are listed in Table~\ref{tab:disk_sources}.

\begin{table}
\centering
\caption{Disk and estimated 100~K radii for sources with both a disk and CH$_3$OH detection}
\label{tab:disk_sources}
\begin{tabular}{llll}
\hline \hline
Source & $R_\mathrm{disk}$ & $R_\mathrm{100K}$ & Normalized $M_\mathrm{CH_3OH}$ \\
 & au & au & \\
\hline
Per-emb~27 & 20\tablefootmark{(1)} & 147 & 6.6$\pm$3.5(-5) \\
Per-emb~35A & 20\tablefootmark{(1)} & 49 & 2.3$\pm$2.2(-5) \\
IRAS~4A1 (VLA) & 35\tablefootmark{(1)} & 46 & $>$4.4(-6) \\
Per-emb~26 & 37\tablefootmark{(2)} & 49 & 1.4$\pm$0.6(-5) \\
Per-emb~1 & 30\tablefootmark{(3)} & 30 & 1.8$\pm$1.7(-6) \\
Per-emb~11A & 40\tablefootmark{(2)} & 30 & 8.8$\pm$8.7(-7) \\
HH~212 & 60\tablefootmark{(4)} & 46 & 8.5$\pm$3.9(-7) \\
L1551~IRS5 & 140\tablefootmark{(5)} & 91 & $>$5.0(-7) \\
V883~Ori & 320\tablefootmark{(6)} & 227 & 1.2$\pm$0.3(-6) \\
\hline
\end{tabular}
\tablefoot{$a(b)$ represents $a\times10^b$. The 100~K radius was determined using the relation derived by \citet{Bisschop2007} for high-mass hot cores, $R_\mathrm{100K} \approx 15.4 \sqrt{L_\mathrm{bol}/L_\odot}$~au, which also seems to hold for low-mass protostars \citep{vantHoff2022}. \\
\tablefoottext{1}{\citet{Segura-Cox2018}.}
\tablefoottext{2}{\citet{Maury2019}.}
\tablefoottext{3}{\citet{Lee2018}.}
\tablefoottext{4}{\citet{Lee2017_HH212_disk}.}
\tablefoottext{5}{\citet{Cruz-SaenzdeMiera2019}.}
\tablefoottext{6}{\citet{Cieza2016}.}
}
\end{table}

The normalized warm $M_\mathrm{CH_3OH}$ of Per-emb~27, Per-emb~35A, and Per-emb~26 lie very well within the 3$\sigma$ range of the best-fit power-law model in Fig.~\ref{fig:M_div_M_dust_vs_L_Diskdetections} (green shaded area). Similarly, the lower limit of IRAS~4A1 (VLA) falls within green area in Fig.~\ref{fig:M_div_M_dust_vs_L_Diskdetections}. This suggests that for these sources the thermal structure of the envelope is not significantly affected by the disk. This suggestion is further supported by the derived radii compared to the estimated $100$~K radius based on their luminosities assuming that the thermal structure of the envelope is not dominated by the disk, see Table~\ref{tab:disk_sources}. Especially for Per-emb~27 (IRAS~2A) the disk is much smaller than the estimated $100$~K radius ($R_\mathrm{disk} = 20$~au and $R_\mathrm{100K} = 147$~au, respectively), suggesting that indeed the disk does not significantly affect the thermal structure. 

On the other hand, V883~Ori shows more than two orders of magnitude lower normalized warm $M_\mathrm{CH_3OH}$ than the 3$\sigma$ range of the best-fit power-law model in Fig.~\ref{fig:M_div_M_dust_vs_L_Diskdetections}. Likewise, HH~212, the ranges of Per-emb~1 and Per-emb~11A, and the lower limit of L1551~IRS5 lie about an order of magnitude below the 3$\sigma$. One possible explanation for this discrepancy is the size of the disk for these sources. The large disks around V883~Ori and L1551~IRS5 compared to their estimated $100$~K radii suggest that these disks have a stronger impact on the thermal structure of these sources compared to the sources discussed above, therefore lowering the amount warm gas in the inner envelope. Since both Per-emb~27 and V883~Ori are currently in an outbursting phase \citep[e.g.,][]{Lee_V883_2019,Hsieh2019}, this cannot explain the more than one order of magnitude difference in normalized $M_\mathrm{CH_3OH}$. For HH~212, the edge-on disk is slightly larger than the 100~K radius, but the COMs in this source are detected through very high-angular resolution observations ($\sim12$~au) in the part of the envelope not shadowed by the disk or the disk atmosphere \citep{Lee2017_HH212,Lee2019_HH212}. The disks around Per-emb~1 and Per-emb~11A are similar in size as the estimated 100~K radius, but since the errorbars on their normalized warm CH$_3$OH masses are rather large no further conclusions can be made.

Of the 46 sources where no disk is confirmed on $>50$~au scales, 26 sources fall within 3$\sigma$ of the best-fit power-law model, indicating that the thermal structure in these sources is indeed not dominated by a (large) disk. However, 20 sources where no disk is confirmed on $>50$~au scales lie up to two orders of magnitude below 3$\sigma$ of the best-fit power-law model. One example is BHR~71, which shows a warm normalized CH$_3$OH mass about an order of magnitude lower than green shaded region in Fig~\ref{fig:M_div_M_dust_vs_L_Diskdetections}. Hints of a $\sim50$~au Keplerian disk are visible in the $^{13}$CH$_3$OH lines \citep{Yang2020}, which is roughly similar in size as the estimated $100$~K radius of $\sim57$~au \citep{Bisschop2007}. Most of the other red upper limits in Fig.~\ref{fig:M_div_M_dust_vs_L_Diskdetections} have constraints on their disk size down to $\sim10$~au scales due to high-angular resolution continuum observation in e.g., the VANDAM survey \citep{Tobin2016,Segura-Cox2018}. Therefore, it is unlikely that large disks are the explanation for the low normalized warm $M_\mathrm{CH_3OH}$ in these sources, but small disks may still be present that can affect the thermal structure. An alternative reason could be the dust opacity (see Sect.~\ref{subsec:cont_tau}).

Among the high-mass sources, only AFGL~4176 is known to host a large $\sim2000$~au disk \citep{Johnston2015}, but it also shows a large amount of COMs including CH$_3$OH in the disk \citep{Bogelund2019}. Several other high-mass sources are likely to also host large warm disks as these seem to be common at least for A and B-type stars \citep{Beltran2016}. However, since no information on the presence or absence of disks is available for the high-mass sources, all other high-mass sources are set in the "other" category.

\subsection{Continuum optical depth}
\label{subsec:cont_tau}
Another way to hide large amounts of COMs is through continuum optical depth at (sub)millimeter wavelengths. If a layer of optically thick dust is present coinciding with or in front of the hot core, all emission from the hot core will be extincted. Alternatively, continuum over-subtraction due to the presence of bright (optically thick) continuum emission behind the methanol emission can hide CH$_3$OH emission \citep[e.g.,][]{Boehler2017,Rosotti2021}. A recent example of a low-mass source where optically thick dust hides emission from COMs at submillimeter wavelengths is NGC~1333~IRAS4A1 \citep{Lopez-Sepulcre2017,deSimone2020}. For high-mass sources, the effect dust opacity on the molecular line intensities of COMs toward G31.41+0.31 was shown by \citet{Rivilla2017}.  Optically thick dust thus does not reduce the amount of COMs present in the gas phase but it hides the emission from the hot core at millimeter wavelengths.

As a zeroth order approximation, the continuum optical depth toward the sources studied in this work can be quantified using the continuum flux and assuming a dust temperature $T_\mathrm{dust}$ \citep[see Eq.~(2) of][]{Rivilla2017},
\begin{align}
\tau_\mathrm{\nu} = -\ln\left(1-\frac{F_\mathrm{\nu}}{\Omega_\mathrm{beam} B_\mathrm{\nu}(T_\mathrm{dust})}\right),
\label{eq:tau_cont}
\end{align}
where $\Omega_\mathrm{beam}$ is the beam solid angle and $F_\nu$ is the flux density within the beam at frequency $\nu$. Here, we assume that $T_\mathrm{dust}=30$~K and $T_\mathrm{dust}=50$~K for the low-mass and high-mass sources, respectively (see Sect.~\ref{subsec:toy_model}). 

For the PEACHES sample, the highest optical depth value at 243~GHz (i.e., 1.23~mm) is for IRAS~4A1 (i.e., Per-emb~12A, $\tau_\mathrm{\nu} = 3.9$) which is consistent with a non-detection of emission by CH$_3$OH \citep[it is in fact in absorption at millimeter wavelengths;][]{Sahu2019} since any COMs emission is extincted with a factor $\approx 50$. The effect of dust opacity on the normalized warm $M_\mathrm{CH_3OH}$ of IRAS~4A1 is also shown in Fig.~\ref{fig:M_div_M_dust_vs_L_Diskdetections} where the lower limit derived at centimeter wavelengths \citep[IRAS~4A1 (VLA);][]{deSimone2020} is more than two orders of magnitude higher than the normalized warm $M_\mathrm{CH_3OH}$ derived from absorption lines at millimeter wavelengths (IRAS~4A1 (ALMA)). In contrast to source A, Per-emb 12B has significantly lower extinction by the dust ($\tau_\mathrm{\nu} = 0.56$), explaining why source B appears rich in COMs at millimeter wavelengths. Moreover, both these optical depth estimates are consistent with those derived by \citet{deSimone2020} at 143~GHz when assuming that $\tau_\nu \propto \nu^{\alpha-2}$ for $\alpha=2.5$ \citep{Tychoniec2020}.

\begin{figure}
\centering
\includegraphics[width=0.9\linewidth]{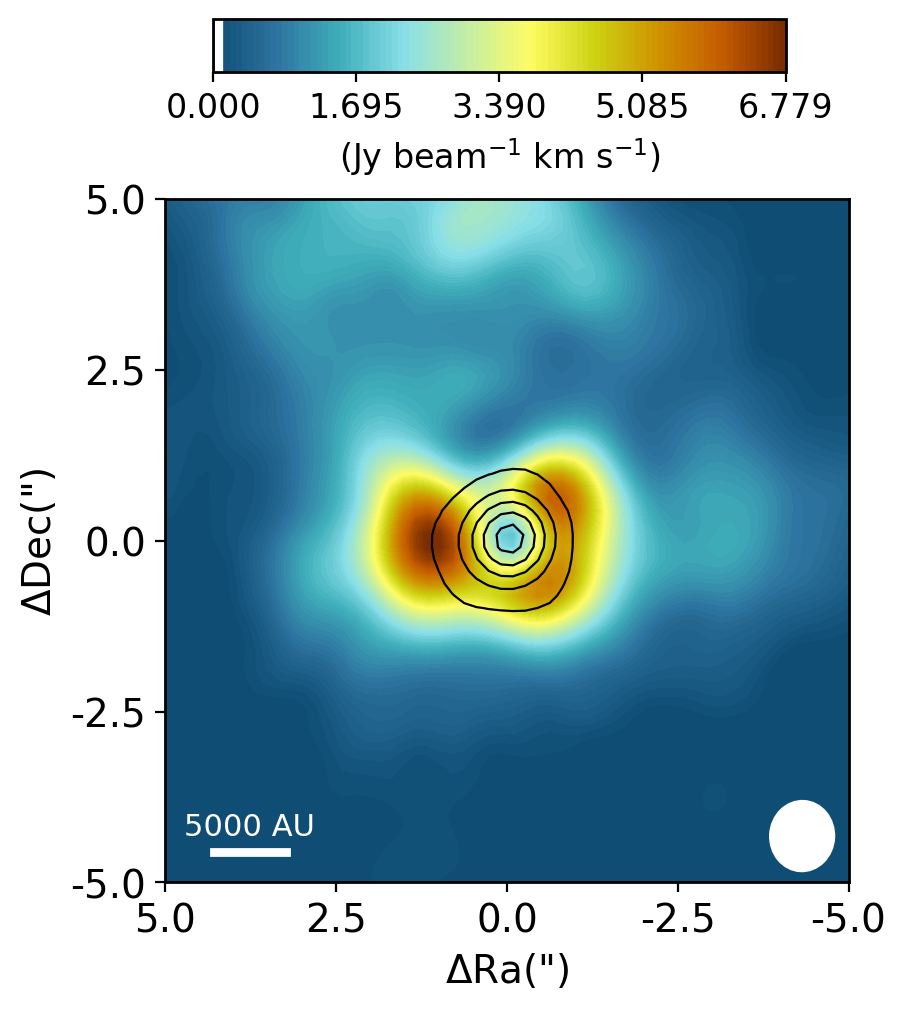}
\caption{
Integrated intensity maps of the CH$_3$OH $8_{0,8}-7_{1,6}$ line for 693050. The color scale is shown on top of the image. The image is integrated over [-5,5]~\kms\ with respect to the $V_\mathrm{lsr}$. The white vertical line in the colorbar indicates the $3\sigma$ threshold. The continuum flux is indicated with the black contours at [0.1,0.3,0.5,0.7,0.9] times the peak continuum flux $F_\mathrm{cont} = 1.92$~Jy~beam$^{-1}$. The white ellipse in the lower right of each image depicts the beam size and in the lower left a physical scale bar is displayed.
}
\label{fig:ALMAGAL_dust_continuum}
\end{figure}

In the ALMAGAL sample, the effect of dust opacity is evidently visible in the source 693050 where the CH$_3$OH emission is ring-shaped around the continuum peak, see Fig.~\ref{fig:ALMAGAL_dust_continuum}. Using Eq.~\eqref{eq:tau_cont} for a dust temperature of 50~K, the continuum optical depth is estimated to be $\tau_\mathrm{\nu} = 0.68$ at 219~GHz. This could explain the decrease in emission in the center, but since the optical depth is not as high as in Per-emb 12A, still CH$_3$OH is detected outside the central beam. Similarly, the CH$_3$OH emission peaks slightly offsource for 707948 ($\tau_\mathrm{\nu} = 0.24$) but still significant emission is present within the central beam. The reason why these optical depth values are lower than for some of the PEACHES sample could be that the solid angle of the optically thick emitting region is in reality much smaller than the beam size. However, the fact that 693050 shows ring-shaped CH$_3$OH emission and also has the highest optical depth of the high-mass sources is a clear case where the dust attenuation is reducing the emission of CH$_3$OH.

\section{Conclusion}
\label{sec:conclusion}
In this work, the warm ($T>100$~K) gaseous methanol mass $M_\mathrm{CH_3OH}$ is derived for a large sample of both low-mass and high-mass embedded protostellar systems. The sample includes new and archival observations of 148 sources with ALMA, as well as 36 sources added from the literature, leading to a combined sample size of 184 sources. Using a simple analytic toy model of a spherically symmetric infalling envelope that is passively heated by the central protostar, the effect of source structure (e.g., a disk) and dust opacity on the amount of warm gaseous methanol is investigated. The main conclusions are as follows:

\begin{itemize}
\item Among the low-mass protostars, a scatter of more than four orders of magnitude in $M_\mathrm{CH_3OH}$ is observed with values ranging between $10^{-7}$~M$_\odot$ and $\lesssim10^{-11}$~M$_\odot$. On average, Class~0 sources have more CH$_3$OH mass ($\sim10^{-7}$~M$_\odot$) than more evolved Class~I sources ($\lesssim10^{-10}$~M$_\odot$). High-mass sources in our biased sample show higher warm gaseous CH$_3$OH masses between $\sim 10^{-7}-10^{-3}$~M$_\odot$. This scatter can be either due to source structure (e.g., presence of a disk) or dust optical depth.

\item To take into account the effect of the source's overall mass and to test the effect of the bolometric luminosity on the snowline radius, the normalized warm $M_\mathrm{CH_3OH}$ is defined as $M_\mathrm{CH_3OH}/M_\mathrm{dust,0}$. Here, $M_\mathrm{dust,0}$ is the cold + warm dust mass in the disk and inner envelope within a fixed radius measured from the ALMA dust continuum. A simple power-law fit to the normalized warm $M_\mathrm{CH_3OH}$ of the line-rich sources gives a positive correlation with the bolometric luminosity: $M_\mathrm{CH_3OH}/M_\mathrm{dust,0} \propto L_\mathrm{bol}^{0.70\pm0.05}$ over an $L_\mathrm{bol}$ range of $10^{-1}-10^{6}$~L$_\odot$. This in good agreement with the toy hot core model which predicts that $M_\mathrm{CH_3OH}/M_\mathrm{0} \propto L_\mathrm{bol}^{0.75}$ and can be considered as an observational confirmation of the scaling between the sublimation radius (i.e., snowline radius) and source luminosity of $R_\mathrm{sub} \propto L_\mathrm{bol}^{1/2}$ found in radiative transfer calculations.

\item Most sources where the disk radius is equivalent to or smaller than the estimated $100$~K envelope radius agree well with the power-law fit to normalized warm $M_\mathrm{CH_3OH}$, indicating that the thermal structure of the envelope in these sources is likely not affected by the disk. On the other hand, sources for which the disk radius is significantly larger have up to two orders of magnitude lower normalized warm CH$_3$OH masses, suggesting that these disks significantly affect the thermal structure of these sources.

\item High dust opacity can hide emission from COMs, leading to low observed CH$_3$OH masses. Clear cases of this effect in our sample are Per-emb~12A (IRAS~4A1) among the low-mass sources and 693050 among the high-mass sources. 
\end{itemize}

\noindent This work shows that the absence of COMs emission in embedded protostellar systems does not mean that the abundance of COMs in such sources is low but that it may be a physical effect due to the structure of the source, most notably the presence of a disk. It is therefore very important to understand the physical structure of embedded protostellar systems in order to understand the chemistry. The modeling work done by \citet{Nazari2022} quantifies the effect that a disk has on the temperature structure of an embedded protostellar system and thus on the amount of warm methanol. Additionally, observations at longer wavelengths can solve the problem of dust optical depth.

\begin{acknowledgements}
The authors would like to thank the anonymous referee for their constructive comments on the manuscript.

This paper makes use of the following ALMA data: ADS/JAO.ALMA\#2017.1.01174.S, ADS/JAO.ALMA\#2017.1.01350.S, ADS/JAO.ALMA\#2016.1.01501.S, ADS/JAO.ALMA\#2017.1.01462.S, and ADS/JAO.ALMA\#2019.1.00195.L. ALMA is a partnership of ESO (representing its member states), NSF (USA) and NINS (Japan), together with NRC (Canada), MOST and ASIAA (Taiwan), and KASI (Republic of Korea), in cooperation with the Republic of Chile. The Joint ALMA Observatory is operated by ESO, AUI/NRAO and NAOJ. 

Astrochemistry in Leiden is supported by the Netherlands Research School for Astronomy (NOVA), by funding from the European Research Council (ERC) under the European Union’s Horizon 2020 research and innovation programme (grant agreement No. 101019751 MOLDISK), and by the Dutch Research Council (NWO) grants TOP-1 614.001.751, 648.000.022, and 618.000.001. Support by the Danish National Research Foundation through the Center of Excellence “InterCat” (Grant agreement no.: DNRF150) is also acknowledged.

G.A.F acknowledges support from the Collaborative Research Centre 956, funded by the Deutsche Forschungsgemeinschaft (DFG) project ID 184018867.

Y.-L. Yang acknowledges the support from the Virginia Initiative of Cosmic Origins (VICO) Postdoctoral Fellowship.
\end{acknowledgements}


\bibliographystyle{aa}
\bibliography{refs}

\clearpage
\onecolumn

\begin{appendix}

\section{Observational details}
\begin{longtable}{lllllcccc}
\caption{Observations of protostars.\label{tab:observations}}\\
\hline\hline
ALMA program & Band & Source & RA (J2000) & Dec (J2000) & $V_\mathrm{lsr}$ & \multicolumn{3}{c}{Detected species} \\  \cline{7-9}
& & & & & \kms & CH$_3$OH & $^{13}$CH$_3$OH & CH$_{3}^{18}$OH\\
\hline
\endfirsthead
\caption{continued.}\\
\hline\hline
ALMA program & Band & Source & RA (J2000) & Dec (J2000) & $V_\mathrm{lsr}$ & \multicolumn{3}{c}{Detected species}\\ \cline{7-9}
& & & & & \kms & CH$_3$OH & $^{13}$CH$_3$OH & CH$_{3}^{18}$OH\\
\hline
\endhead
\hline
\endfoot
2017.1.01174.S  & 6 & B1-c &  03:33:17.88 & 31:09:31.8 & 6.0 & Y & Y & Y \\
                &       & B1-b & 03:33:20.34 & 31:07:21.3 & 6.0 & N & N & N \\
                                &       & B1-bN & 03:33:21.21 & 31:07:43.6 & 6.0 & N & N & N \\
                                &       & B1-bS & 03:33:21.36 & 31:07:26.3 & 6.0 & Y & Y & Y \\
                                &       & Serpens S68N & 18:29:48.09 & 01:16:43.3 & 8.5 & Y & Y & Y \\
                                &       & Ser-emb 8 (N) & 18:29:48.73 & 01:16:55.6 & 8.5 & N & N & N \\
\hline
2017.1.01350.S  & 6 & Serpens SMM3 & 18:29:57.75 & 01:14:06.7 & 8.5 & N & N & N \\
\hline
2016.1.01501.S  & 6 & Per-emb 22B & 03:25:22.35 & 30:45:13.11 & 4.3 & Y & N & N \\
				& 	& Per-emb 22A & 03:25:22.41 & 30:45:13.26 & 4.3 & Y & Y & Y \\
				& 	& L1448 NW & 03:25:35.67 & 30:45:34.16 & 4.2 & N & N & N \\
				& 	& Per-emb 33BC & 03:25:36.32 & 30:45:15.19 & 5.3 & N & N & N \\
				& 	& Per-emb 33A & 03:25:36.38 & 30:45:14.72 & 4.3 & Y & N & N \\
				& 	& L1448 IRS 3A & 03:25:36.50 & 30:45:21.90 & 4.6 & Y & N & N \\
				& 	& Per-emb 26 & 03:25:38.88 & 30:44:05.28 & 5.4 & Y & Y & N \\
				& 	& Per-emb 42 & 03:25:39.14 & 30:43:57.90 & 5.8 & Y & N & N \\
				& 	& Per-emb 25 & 03:26:37.51 & 30:15:27.81 & 5.5 & N & N & N \\
				& 	& Per-emb 17 & 03:27:39.11 & 30:13:02.96 & 6.0 & Y & Y & Y \\
				& 	& Per-emb 20 & 03:27:43.28 & 30:12:28.88 & 5.3 & Y & N & N \\
				& 	& L1455 IRS 2 & 03:27:47.69 & 30:12:04.33 & 5.1 & N & N & N \\
				& 	& Per-emb 35A & 03:28:37.10 & 31:13:30.77 & 7.4 & Y & N & N \\
				& 	& Per-emb 35B & 03:28:37.22 & 31:13:31.74 & 7.3 & Y & N & N \\
				& 	& Per-emb 27 & 03:28:55.57 & 31:14:36.97 & 6.5 & Y & Y & Y \\
				& 	& EDJ2009 172 & 03:28:56.65 & 31:18:35.43 & -- & N & N & N \\
				& 	& Per-emb 36 & 03:28:57.37 & 31:14:15.77 & 6.9 & N & N & N \\
				& 	& Per-emb 54 & 03:29:01.55 & 31:20:20.49 & 7.9 & N & N & N \\
				& 	& SVS 13B & 03:29:03.08 & 31:15:51.73 & 8.5 & N & N & N \\
				& 	& SVS 13A2 & 03:29:03.39 & 31:16:01.58 & 8.4 & Y & N & N \\
				& 	& Per-emb 44 & 03:29:03.76 & 31:16:03.70 & 8.7 & Y & Y & Y \\
				& 	& Per-emb 15 & 03:29:04.06 & 31:14:46.23 & 6.8 & N & N & N \\
				& 	& Per-emb 50 & 03:29:07.77 & 31:21:57.11 & 9.3 & N & N & N \\
				& 	& Per-emb 12B & 03:29:10.44 & 31:13:32.08 & 6.9 & Y & Y & Y \\
				& 	& Per-emb 12A & 03:29:10.54 & 31:13:30.93 & 6.9 & A\tablefootmark{(1)} & N & N \\
				& 	& Per-emb 21 & 03:29:10.67 & 31:18:20.16 & 8.6 & N & N & N \\
				& 	& Per-emb 18 & 03:29:11.27 & 31:18:31.09 & 8.1 & Y & N & N \\
				& 	& Per-emb 13 & 03:29:12.02 & 31:13:07.99 & 7.1 & Y & Y & Y \\
				& 	& IRAS 4B2 & 03:29:12.85 & 31:13:06.87 & 7.1 & N & N & N \\
				& 	& Per-emb 14 & 03:29:13.55 & 31:13:58.12 & 7.9 & N & N & N \\
				& 	& EDJ2009 235 & 03:29:18.26 & 31:23:19.73 & 7.7 & N & N & N \\
				& 	& EDJ2009 237 & 03:29:18.74 & 31:23:25.24 & -- & N & N & N \\
				& 	& Per-emb 37 & 03:29:18.97 & 31:23:14.28 & 7.5 & N & N & N \\
				& 	& Per-emb 60 & 03:29:20.05 & 31:24:07.35 & -- & N & N & N \\
\hline
2017.1.01462.S  & 6	& Per-emb 5 & 03:31:20.94 & 30:45:30.24 & 7.3 & Y & Y & N \\
				& 	& Per-emb 2 & 03:32:17.92 & 30:49:47.81 & 7.0 & N & N & N \\
				& 	& Per-emb 10 & 03:33:16.43 & 31:06:52.01 & 6.4 & Y & N & N \\
				& 	& Per-emb 40 & 03:33:16.67 & 31:07:54.87 & 7.4 & N & N & N \\
				& 	& Per-emb 29 & 03:33:17.88 & 31:09:31.74 & 6.1 & Y & Y & Y \\
				& 	& B1-bN & 03:33:21.21 & 31:07:43.63 & 6.6 & N & N & N \\
				& 	& B1-bS & 03:33:21.36 & 31:07:26.34 & 6.6 & Y & Y & N \\
				& 	& Per-emb 16 & 03:43:50.97 & 32:03:24.12 & 8.8 & N & N & N \\
				& 	& Per-emb 28 & 03:43:51.01 & 32:03:08.02 & 8.6 & N & N & N \\
				& 	& Per-emb 1 & 03:43:56.81 & 32:00:50.16 & 9.4 & Y & N & N \\
				& 	& Per-emb 11B & 03:43:56.88 & 32:03:03.08 & 9.0 & N & N & N \\
				& 	& Per-emb 11A & 03:43:57.07 & 32:03:04.76 & 9.0 & Y & N & N \\
				& 	& Per-emb 11C & 03:43:57.70 & 32:03:09.82 & 9.0 & N & N & N \\
				& 	& Per-emb 55 & 03:44:43.30 & 32:01:31.22 & 12.0 & N & N & N \\
				& 	& Per-emb 8 & 03:44:43.98 & 32:01:35.19 & 11.0 & N & N & N \\
				& 	& Per-emb 53 & 03:47:41.59 & 32:51:43.62 & 10.2 & N & N & N \\
\hline
2019.1.00195.L	& 6	& 81635A & 18:25:00.82 & -13:15:34.46 & -- & N & N & N \\
				&	& 81635B & 18:25:01.01 & -13:15:38.57 & -- & N & N & N \\
				&	& 81635C & 18:25:01.65 & -13:15:28.99 & -- & N & N & N \\
				&	& 83968A & 18:25:10.57 & -12:42:22.43 & -- & N & N & N \\
				&	& 83968B & 18:25:10.69 & -12:42:26.14 & -- & N & N & N \\
				&	& 83968C & 18:25:10.80 & -12:42:24.68 & 46.2 & Y & N & N \\
				&	& 83968D & 18:25:10.57 & -12:42:19.91 & -- & N & N & N \\
				&	& 83968E & 18:25:10.64 & -12:42:24.53 & -- & N & N & N \\
				&	& 86213A & 18:26:48.89 & -12:26:23.75 & -- & N & N & N \\
				&	& 86213B & 18:26:47.94 & -12:26:20.77 & 66.0 & Y & N & N \\
				&	& 86213C & 18:26:48.73 & -12:26:26.24 & 63.9 & Y & N & N \\
				&	& 101899 & 18:34:40.28 & -09:00:38.33 & 77.3 & Y & Y & N \\
				&	& 103421 & 18:33:23.96 & -08:33:31.64 & 78.5 & Y & Y & N \\
				&	& 106756A & 18:34:23.99 & -07:54:48.28 & 80.0 & Y & N & N \\
				&	& 106756B & 18:34:25.55 & -07:54:46.39 & -- & N & N & N \\
				&	& 106756C & 18:34:25.59 & -07:54:43.11 & -- & N & N & N \\
				&	& 126120A & 18:42:37.58 & -04:02:05.58 & 80.2 & Y & N & N \\
				&	& 126120B & 18:42:37.64 & -04:02:07.32 & 81.2 & Y & N & N \\
				&	& 126120C & 18:42:36.85 & -04:02:17.66 & -- & N & N & N \\
				&	& 126120D & 18:42:37.14 & -04:02:02.37 & -- & N & N & N \\
				&	& 126348 & 18:42:51.99 & -03:59:54.06 & 76.9 & Y & Y & Y \\
				&	& 565926A & 08:02:42.97 & -34:31:48.77 & -- & N & N & N \\
				&	& 565926B & 08:02:42.94 & -34:31:49.96 & -- & N & N & N \\
				&	& 565926C & 08:02:42.72 & -34:31:49.61 & -- & N & N & N \\
				&	& 586092A & 08:32:08.65 & -43:13:45.58 & 9.7 & Y & Y & N \\
				&	& 586092B & 08:32:08.47 & -43:13:49.06 & 15.5 & Y & N & N \\
				&	& 586092C & 08:32:09.06 & -43:13:43.28 & -- & N & N & N \\
				&	& 615590 & 09:24:41.96 & -52:02:07.98 & 40.4 & Y & Y & Y \\
				&	& 640076 & 10:20:15.74 & -58:03:55.86 & 10.9 & Y & N & N \\
				&	& 644284A & 10:31:29.73 & -58:02:19.54 & 0.5 & Y & Y & Y \\
				&	& 644284B & 10:31:29.60 & -58:02:18.57 & 2.7 & Y & Y & Y \\
				&	& 693050 & 12:35:35.09 & -63:02:31.97 & -41.0 & Y & Y & Y \\
				&	& 695243 & 12:43:31.51 & -62:36:13.25 & -- & N & N & N \\
				&	& 704792 & 13:11:14.14 & -62:45:06.80 & -- & N & N & N \\
				&	& 705768 & 13:12:36.18 & -62:33:34.75 & -34.5 & Y & Y & Y \\
				&	& 706733A & 13:14:22.77 & -62:45:59.07 & -- & N & N & N \\
				&	& 706733B & 13:14:22.99 & -62:45:54.35 & -- & N & N & N \\
				&	& 706733C & 13:14:23.07 & -62:45:47.54 & -- & N & N & N \\
				&	& 706785A & 13:14:26.86 & -62:44:29.95 & -39.8 & Y & N & N \\
				&	& 706785B & 13:14:26.52 & -62:44:31.40 & -40.8 & Y & N & N \\
				&	& 706785C & 13:14:26.38 & -62:44:30.24 & -- & N & N & N \\
				&	& 706785D & 13:14:25.68 & -62:44:29.97 & -40.2 & Y & N & N \\
				&	& 707948 & 13:16:43.21 & -62:58:32.86 & -35.4 & Y & Y & Y \\
				&	& 717461A & 13:43:01.66 & -62:08:51.37 & -53.5 & Y & Y & Y \\
				&	& 717461B & 13:43:01.74 & -62:08:55.34 & -- & N & N & N \\
				&	& 721992 & 13:51:58.32 & -61:15:41.19 & -57.5 & Y & N & N \\
				&	& 724566 & 13:59:30.93 & -61:48:38.20 & -57.3 & Y & Y & N \\
				&	& 732038 & 14:13:15.05 & -61:16:52.92 & -63.4 & Y & N & N \\
				&	& 744757A & 14:45:26.37 & -59:49:15.29 & -41.1 & Y & Y & Y \\
				&	& 744757B & 14:45:26.13 & -59:49:19.56 & -41.0 & Y & N & N \\
				&	& 759150A & 15:10:43.04 & -57:44:49.11 & -- & N & N & N \\
				&	& 759150B & 15:10:43.52 & -57:44:44.82 & -- & N & N & N \\
				&	& 759150C & 15:10:44.48 & -57:44:47.33 & -- & N & N & N \\
				&	& 759150D & 15:10:42.71 & -57:44:52.85 & -- & N & N & N \\
				&	& 759150E & 15:10:44.10 & -57:44:52.03 & -- & N & N & N \\
				&	& 767784 & 15:29:19.37 & -56:31:22.25 & -68.4 & Y & Y & Y \\
				&	& 800287 & 16:11:26.56 & -51:41:56.99 & -83.5& Y & Y & N \\
				&	& 854214A & 16:52:32.73 & -43:23:49.20 & -71.0 & Y & N & N \\
				&	& 854214B & 16:52:33.14 & -43:23:49.42 & -72.2 & Y & N & N \\
				&	& 863312A & 17:02:08.36 & -41:46:56.89 & -- & N & N & N \\
				&	& 863312B & 17:02:09.14 & -41:46:45.04 & -- & N & N & N \\
				&	& 865468A & 17:05:10.90 & -41:29:06.93 & -26.8 & Y & Y & N \\
				&	& 865468B & 17:05:11.20 & -41:29:07.21 & -27.5 & Y & Y & Y \\
				&	& 865468C & 17:05:11.10 & -41:29:03.22 & -25.7 & Y & Y & Y \\
				&	& 876288 & 17:11:51.03 & -39:09:29.12 & -96.0 & Y & Y & Y \\
				&	& 881427A & 17:20:06.29 & -38:57:14.88 & -12.0 & Y & Y & Y \\
				&	& 881427B & 17:20:06.42 & -38:57:11.24 & -10.6 & Y & Y & Y \\
				&	& 881427C & 17:20:06.10 & -38:57:15.81 & -10.4 & Y & Y & Y \\
				&	& G023.3891+00.1851 & 18:33:14.32 & -08:23:57.61 & 75.5 & Y & Y & N \\
				&	& G023.6566-00.1273 & 18:34:51.55 & -08:18:21.57 & 81.2 & Y & Y & N \\
				&	& G025.6498+01.0491 & 18:34:20.92 & -05:59:42.02 & 41.2 & Y & Y & Y \\
				&	& G030.1981-00.1691 & 18:47:03.06 & -02:30:36.39 & 105.3 & Y & Y & N \\
				&	& G233.8306-00.1803 & 07:30:16.73 & -18:35:49.06 & -- & N & N & N \\
				&	& G305.2017+00.2072A1 & 13:11:10.47 & -62:34:38.58 & -41.2 & Y & Y & Y \\
				&	& G305.2017+00.2072A2 & 13:11:13.12 & -62:34:42.74 & -40.9 & Y & N & N \\
				&	& G310.0135+00.3892 & 13:51:37.84 & -61:39:07.53 & -43.0 & Y & N & N \\
				&	& G314.3197+00.1125 & 14:26:26.26 & -60:38:31.34 & -50.0 & Y & N & Y \\
				&	& G316.6412-00.0867 & 14:44:18.33 & -59:55:11.52 & -20.0 & Y & Y & Y \\
				&	& G318.0489+00.0854B & 14:53:42.68 & -59:08:53.01 & -50.4 & Y & Y & Y \\
				&	& G318.9480-00.1969A1 & 15:00:55.30 & -58:58:52.42 & -35.0 & Y & Y & Y \\
				&	& G318.9480-00.1969A2 & 15:00:55.23 & -58:58:55.88 & -35.0 & Y & N & N \\
				&	& G323.7399-00.2617B1 & 15:31:45.64 & -56:30:50.16 & -51.5 & Y & Y & Y \\
				&	& G323.7399-00.2617B2 & 15:31:45.45 & -56:30:50.06 & -51.3 & Y & Y & Y \\
				&	& G323.7399-00.2617B3 & 15:31:45.73 & -56:30:51.93 & -51.3 & Y & N & N \\
				&	& G323.7399-00.2617B4 & 15:31:45.94 & -56:30:51.34 & -50.5 & Y & N & N \\
				&	& G323.7399-00.2617B5 & 15:31:45.62 & -56:30:45.62 & -52.0 & Y & N & N \\
				&	& G323.7399-00.2617B6 & 15:31:45.84 & -56:30:47.68 & -49.7 & Y & N & N \\
				&	& G323.7399-00.2617B7 & 15:31:45.91 & -56:30:46.10 & -49.3 & Y & N & N \\
				&	& G327.1192+00.5103 & 15:47:32.72 & -53:52:38.55 & -84.7 & Y & Y & Y \\
				&	& G343.1261-00.0623 & 16:58:17.23 & -42:52:07.67 & -35.0 & Y & Y & N \\
				&	& G345.5043+00.3480 & 17:04:22.88 & -40:44:22.92 & -19.7 & Y\tablefootmark{(2)} & Y\tablefootmark{(2)} & Y\tablefootmark{(2)} \\
				&	& G348.7342-01.0359B1 & 17:20:07.12 & -38:57:11.69 & -10.7 & Y & N & N \\
				&	& G348.7342-01.0359B2 & 17:20:07.26 & -38:57:09.82 & -13.1 & Y & N & N \\
				&	& G348.7342-01.0359B3 & 17:20:07.38 & -38:57:10.15 & -- & N & N & N \\
\end{longtable}
\tablefoot{The coordinates mark the peak of the continuum emission. The last three columns indicate if unblended transitions of CH$_3$OH and its isotopologues are detected at the $3\sigma$ limit (Y) or if the all the lines of a particular isotopologue were either too blended for a fit or not detected at the $3\sigma$ detection (N). \\
\tablefoottext{1}{Lines in absorption}
\tablefoottext{2}{Lines consist of two components. Fits are done to the strongest peak.}}

\begin{landscape}
\begin{longtable}{lllllllllllllll}
\caption{Protostellar properties and derived physical quantities.\label{tab:properties_quantities}}\\
\hline\hline
Source & Class & Disk & $d$ & $L_\mathrm{bol}$ &  
$F_\mathrm{cont}$ & 
$\theta_\mathrm{beam,cont}$ & 
$\lambda_\mathrm{cont}$ & $M_\mathrm{dust,0}$ & $\theta_\mathrm{beam}$ & $\theta_\mathrm{source}$ & $N_\mathrm{CH_3OH}$ & 
$M_\mathrm{CH_3OH}$
& Ref.
\\
 & & & pc & L$_\odot$ & 
mJy/beam &
$^{\prime\prime}$ & 
mm & M$_\odot$ & $^{\prime\prime}$ & $^{\prime\prime}$ & cm$^{-2}$ & 
M$_\odot$ 
&
\\
\hline
\endfirsthead
\caption{continued.}\\
\hline\hline
Source & Class & Disk & $d$ & $L_\mathrm{bol}$ & 
$F_\mathrm{cont}$ & 
$\theta_\mathrm{beam,cont}$ &
$\lambda_\mathrm{cont}$ & $M_\mathrm{dust,0}$ & $\theta_\mathrm{beam}$ & $\theta_\mathrm{source}$ & $N_\mathrm{CH_3OH}$ & 
$M_\mathrm{CH_3OH}$
& Ref.
\\
 & & & pc & L$_\odot$ & 
mJy/beam &
$^{\prime\prime}$ & 
mm & M$_\odot$ & $^{\prime\prime}$ & $^{\prime\prime}$ & cm$^{-2}$ & 
M$_\odot$
&
\\
\hline
\endhead
\hline
\endfoot
B1-c & 0 & N & 320 & 5.9 & 93 & 0.45 & 1.14 & 4.0$\pm$0.6(-3) & 0.45 & 0.45 &  1.9$\pm$0.6(18) &  4.6$\pm$1.5(-8) & TW,1,2,3 \\
Serpens S68N & 0 & N & 436 & 5.4 & 44 & 0.45 & 1.14 & 2.2$\pm$0.3(-3) & 0.45 & 0.45 &  1.4$\pm$0.6(18) &  6.3$\pm$2.7(-8) & TW,3,4,5 \\
B1-bS & 0 & N & 320 & 0.6 & 1.5(2) & 0.39 & 1.22 & 9.2$\pm$1.4(-3) & 0.45 & 0.45 &  5.0$\pm$0.6(17) &  1.2$\pm$0.2(-8) & TW,1,3,6 \\
B1-b & I & N & 320 & 0.3 & 9.9 & 0.45 & 1.14 & 4.2$\pm$0.6(-4) & 0.45 & 0.45 & $<$1.0(15) & $<$2.4(-11) & TW,1,3,7 \\
B1-bN & 0 & N & 320 & 0.3 & 76 & 0.39 & 1.22 & 4.8$\pm$0.7(-3) & 0.45 & 0.45 & $<$1.0(15) & $<$2.4(-11) & TW,1,3,6 \\
\hline
Serpens SMM3 & 0 & -- & 436 & 28 & 48 & 0.45 & 1.33 & 3.5$\pm$0.5(-3) & 0.45 & 0.45 & $<$1.0(15) & $<$4.5(-11) & TW,2,3,4 \\
\hline
B1-bN & 0 & N & 320 & 0.3 & 76 & 0.39 & 1.22 & 4.8$\pm$0.7(-3) & 0.39 & 0.39 & $<$6.3(15) & $<$1.2(-10) & TW,1,6 \\
B1-bS & 0 & N & 320 & 0.6 & 1.5(2) & 0.39 & 1.22 & 9.2$\pm$1.4(-3) & 0.39 & 0.39 &  4.1$\pm$1.4(17) &  7.6$\pm$2.5(-9) & TW,1,6 \\
EDJ2009 172 & II & N & 320 & 0.8 & 12 & 0.58 & 1.22 & 4.1$\pm$0.6(-4) & 0.58 & 0.58 & $<$1.4(15) & $<$5.7(-11) & TW,1,8 \\
EDJ2009 235 & II & N & 320 & 4.0(-2) & 5.5 & 0.58 & 1.22 & 1.9$\pm$0.3(-4) & 0.58 & 0.58 & $<$1.7(15) & $<$6.7(-11) & TW,1,6 \\
EDJ2009 237 & II & N & 320 & -- & 2.3 & 0.58 & 1.22 & 8.1$\pm$1.2(-5) & 0.58 & 0.58 & $<$1.9(15) & $<$7.7(-11) & TW,1 \\
IRAS 4B2 & 0 & N & 320 & 3.4 & 1.2(2) & 0.58 & 1.22 & 4.0$\pm$0.6(-3) & 0.58 & 0.58 & $<$6.6(15) & $<$2.7(-10) & TW,1,6 \\
L1448 IRS 3A & I & N & 320 & 18 & 48 & 0.53 & 1.22 & 1.9$\pm$0.3(-3) & 0.53 & 0.53 &  1.4$\pm$1.3(17) &  4.7$\pm$4.6(-9) & TW,1,8 \\
L1448 NW & 0 & N & 320 & 2.7 & 50 & 0.54 & 1.22 & 1.9$\pm$0.3(-3) & 0.54 & 0.54 & $<$2.6(15) & $<$9.3(-11) & TW,1,8 \\
L1455 IRS 2 & Flat & N & 320 & 4.8 & 2 & 0.53 & 1.22 & 8.0$\pm$1.2(-5) & 0.53 & 0.53 & $<$2.4(15) & $<$8.2(-11) & TW,1,8 \\
Per-emb 1 & 0 & Y & 320 & 3.5 & 52 & 0.38 & 1.22 & 3.4$\pm$0.5(-3) & 0.38 & 0.38 &  3.5$\pm$3.4(17) &  6.0$\pm$5.9(-9) & TW,1,8,9 \\
Per-emb 2 & 0 & N & 320 & 1.7 & 66 & 0.39 & 1.22 & 4.1$\pm$0.6(-3) & 0.39 & 0.39 & $<$4.5(15) & $<$8.2(-11) & TW,1,8 \\
Per-emb 5 & 0 & N & 320 & 2.5 & 1.5(2) & 0.39 & 1.22 & 9.2$\pm$1.4(-3) & 0.39 & 0.39 &  1.4$\pm$0.7(18) &  2.5$\pm$1.3(-8) & TW,1,8 \\
Per-emb 8 & 0 & Y & 320 & 3.6 & 81 & 0.38 & 1.22 & 5.3$\pm$0.8(-3) & 0.38 & 0.38 & $<$6.2(15) & $<$1.1(-10) & TW,1,6,9 \\
Per-emb 10 & 0 & N & 320 & 0.8 & 17 & 0.38 & 1.22 & 1.1$\pm$0.2(-3) & 0.38 & 0.38 &  1.4$\pm$1.3(17) &  2.4$\pm$2.3(-9) & TW,1,6 \\
Per-emb 11A & 0 & Y & 320 & 4.3 & 1.1(2) & 0.38 & 1.22 & 6.9$\pm$1.0(-3) & 0.38 & 0.38 &  3.5$\pm$3.4(17) &  6.0$\pm$5.9(-9) & TW,1,8,9 \\
Per-emb 11B & 0 & N & 320 & 8.7(-2) & 4.5 & 0.38 & 1.22 & 3.0$\pm$0.4(-4) & 0.38 & 0.38 & $<$3.8(15) & $<$6.6(-11) & TW,1,8 \\
Per-emb 11C & 0 & N & 320 & 0.1 & 5.3 & 0.38 & 1.22 & 3.5$\pm$0.5(-4) & 0.38 & 0.38 & $<$4.0(15) & $<$7.0(-11) & TW,1,8 \\
Per-emb 12A & 0 & Y & 320 & 9.1 & 3.9(2) & 0.59 & 1.22 & 1.3$\pm$0.2(-2) & 0.59 & 0.59 &  1.4$\pm$1.3(17) &  5.8$\pm$5.7(-9) & TW,1,8,9 \\
Per-emb 12B & 0 & N & 320 & 4.5 & 1.9(2) & 0.59 & 1.22 & 6.4$\pm$1.0(-3) & 0.59 & 0.59 &  6.4$\pm$4.5(18) &  2.7$\pm$1.9(-7) & TW,1,8 \\
Per-emb 13 & 0 & N & 320 & 8.6 & 2.3(2) & 0.58 & 1.22 & 8.1$\pm$1.2(-3) & 0.58 & 0.58 &  2.9$\pm$0.8(18) &  1.2$\pm$0.3(-7) & TW,1,6 \\
Per-emb 14 & 0 & Y & 320 & 1.4 & 55 & 0.58 & 1.22 & 1.9$\pm$0.3(-3) & 0.58 & 0.58 & $<$1.7(15) & $<$7.0(-11) & TW,1,8,9 \\
Per-emb 15 & 0 & N & 320 & 0.8 & 4.1 & 0.58 & 1.22 & 1.4$\pm$0.2(-4) & 0.58 & 0.58 & $<$1.7(15) & $<$6.8(-11) & TW,1,8 \\
Per-emb 16 & 0 & N & 320 & 0.8 & 3.9 & 0.39 & 1.22 & 2.4$\pm$0.4(-4) & 0.39 & 0.39 & $<$3.9(15) & $<$7.1(-11) & TW,1,8 \\
Per-emb 17 & 0 & N & 320 & 8.1 & 30 & 0.53 & 1.22 & 1.2$\pm$0.2(-3) & 0.53 & 0.53 &  1.1$\pm$0.7(18) &  3.7$\pm$2.2(-8) & TW,1,8 \\
Per-emb 18 & 0 & N & 320 & 9.1 & 44 & 0.58 & 1.22 & 1.5$\pm$0.2(-3) & 0.58 & 0.58 &  7.7$\pm$7.6(17) &  3.1$\pm$3.0(-8) & TW,1,6 \\
Per-emb 20 & 0 & N & 320 & 2.7 & 2.6 & 0.53 & 1.22 & 1.0$\pm$0.2(-4) & 0.53 & 0.53 &  3.5$\pm$3.4(17) &  1.2$\pm$1.1(-8) & TW,1,8 \\
Per-emb 21 & 0 & N & 320 & 13 & 39 & 0.58 & 1.22 & 1.4$\pm$0.2(-3) & 0.58 & 0.58 & $<$1.2(15) & $<$4.7(-11) & TW,1,8 \\
Per-emb 22A & 0 & N & 320 & 6.1 & 27 & 0.53 & 1.22 & 1.1$\pm$0.2(-3) & 0.53 & 0.53 &  1.7$\pm$1.1(18) &  5.7$\pm$3.8(-8) & TW,1,6 \\
Per-emb 22B & 0 & N & 320 & 2.2 & 10 & 0.53 & 1.22 & 4.0$\pm$0.6(-4) & 0.53 & 0.53 &  2.1$\pm$2.0(17) &  7.0$\pm$6.8(-9) & TW,1,6 \\
Per-emb 25 & 0 & Y & 320 & 2.3 & 82 & 0.53 & 1.22 & 3.2$\pm$0.5(-3) & 0.53 & 0.53 & $<$3.0(15) & $<$1.0(-10) & TW,1,8,9 \\
Per-emb 26 & 0 & Y & 320 & 10 & 1.3(2) & 0.53 & 1.22 & 5.1$\pm$0.8(-3) & 0.53 & 0.53 &  2.1$\pm$0.9(18) &  7.0$\pm$3.1(-8) & TW,1,6,10 \\
Per-emb 27 & Burst & Y & 320 & 91 & 1.0(2) & 0.59 & 1.22 & 3.5$\pm$0.5(-3) & 0.59 & 0.59 &  5.6$\pm$2.8(18) &  2.3$\pm$1.2(-7) & TW,1,6,9 \\
Per-emb 28 & 0 & N & 320 & 1.4 & 14 & 0.39 & 1.22 & 8.8$\pm$1.3(-4) & 0.39 & 0.39 & $<$9.3(15) & $<$1.7(-10) & TW,1,8 \\
Per-emb 29 & 0 & N & 320 & 5.9 & 86 & 0.39 & 1.22 & 5.4$\pm$0.8(-3) & 0.39 & 0.39 &  8.4$\pm$5.6(18) &  1.5$\pm$1.0(-7) & TW,1,2 \\
Per-emb 33A & 0 & N & 320 & 16 & 1.4(2) & 0.53 & 1.22 & 5.6$\pm$0.8(-3) & 0.53 & 0.53 &  1.0$\pm$0.9(17) &  3.5$\pm$3.4(-9) & TW,1,6 \\
Per-emb 33BC & 0 & N & 320 & 4.8 & 65 & 0.53 & 1.22 & 2.6$\pm$0.4(-3) & 0.53 & 0.53 & $<$2.8(15) & $<$9.5(-11) & TW,1,6 \\
Per-emb 35A & I & Y & 320 & 10 & 17 & 0.58 & 1.22 & 5.9$\pm$0.9(-4) & 0.58 & 0.58 &  3.3$\pm$3.1(17) &  1.3$\pm$1.2(-8) & TW,1,8,9 \\
Per-emb 35B & I & N & 320 & 7.4 & 12 & 0.58 & 1.22 & 4.3$\pm$0.6(-4) & 0.58 & 0.58 &  1.1$\pm$1.0(17) &  4.5$\pm$4.4(-9) & TW,1,8 \\
Per-emb 36 & I & N & 320 & 10 & 1.0(2) & 0.58 & 1.22 & 3.6$\pm$0.5(-3) & 0.58 & 0.58 & $<$2.3(15) & $<$9.4(-11) & TW,1,6 \\
Per-emb 37 & 0 & N & 320 & 0.9 & 8.9 & 0.58 & 1.22 & 3.1$\pm$0.5(-4) & 0.58 & 0.58 & $<$1.8(15) & $<$7.2(-11) & TW,1,6 \\
Per-emb 40 & I & N & 320 & 2.9 & 13 & 0.38 & 1.22 & 8.5$\pm$1.3(-4) & 0.38 & 0.38 & $<$4.2(15) & $<$7.3(-11)  & TW,1,8\\
Per-emb 42 & I & N & 320 & 3.8 & 9.7 & 0.53 & 1.22 & 3.8$\pm$0.6(-4) & 0.53 & 0.53 &  1.9$\pm$1.8(17) &  6.5$\pm$6.4(-9) & TW,1,6 \\
Per-emb 44 & Burst & N & 320 & 57 & 1.1(2) & 0.58 & 1.22 & 3.8$\pm$0.6(-3) & 0.58 & 0.58 &  1.1$\pm$0.6(19) &  4.5$\pm$2.3(-7) & TW,1,8 \\
Per-emb 50 & I & Y & 320 & 45 & 76 & 0.58 & 1.22 & 2.6$\pm$0.4(-3) & 0.58 & 0.58 & $<$1.7(15) & $<$6.8(-11) & TW,1,8,9 \\
Per-emb 53 & I & N & 320 & 9.1 & 15 & 0.38 & 1.22 & 9.6$\pm$1.4(-4) & 0.38 & 0.38 & $<$4.2(15) & $<$7.2(-11) & TW,1,6 \\
Per-emb 54 & I & N & 320 & 32 & 1.8 & 0.58 & 1.22 & 6.1$\pm$0.9(-5) & 0.58 & 0.58 & $<$1.6(15) & $<$6.5(-11) & TW,1,6 \\
Per-emb 55 & I & N & 320 & 2.9 & 3.1 & 0.38 & 1.22 & 2.0$\pm$0.3(-4) & 0.38 & 0.38 & $<$4.2(15) & $<$7.4(-11) & TW,1,6 \\
Per-emb 60 & I & N & 320 & 0.5 & 1.3 & 0.58 & 1.22 & 4.4$\pm$0.7(-5) & 0.58 & 0.58 & $<$1.4(15) & $<$5.6(-11) & TW,1,8 \\
SVS 13A2 & I & N & 320 & 6.3 & 11 & 0.58 & 1.22 & 3.9$\pm$0.6(-4) & 0.58 & 0.58 &  1.7$\pm$1.6(17) &  7.0$\pm$6.9(-9) & TW,1,8 \\
SVS 13B & 0 & Y & 320 & 1.9 & 1.0(2) & 0.58 & 1.22 & 3.6$\pm$0.5(-3) & 0.58 & 0.58 & $<$1.7(15) & $<$6.8(-11) & TW,1,8,9 \\
\hline
81635A & HM & -- & 4.0(3) & 2.4(3) & 39 & 1.22 & 1.37 & 1.4$\pm$0.2(-3) & 1.22 & 1.22 & $<$2.1(15) & $<$5.8(-8) & TW,11,12 \\
81635B & HM & -- & 4.0(3) & 1.4(3) & 24 & 1.22 & 1.37 & 8.8$\pm$1.3(-4) & 1.22 & 1.22 & $<$1.7(15) & $<$4.6(-8) & TW,11,12 \\
81635C & HM & -- & 4.0(3) & 1.3(3) & 21 & 1.22 & 1.37 & 7.7$\pm$1.1(-4) & 1.22 & 1.22 & $<$1.9(15) & $<$5.1(-8) & TW,11,12 \\
83968A & HM & -- & 3.4(3) & 3.2(3) & 24 & 1.23 & 1.37 & 8.1$\pm$1.2(-4) & 1.23 & 1.23 & $<$1.1(15) & $<$2.3(-8) & TW,11,12 \\
83968B & HM & -- & 3.4(3) & 2.0(3) & 15 & 1.23 & 1.37 & 5.1$\pm$0.8(-4) & 1.23 & 1.23 & $<$1.3(15) & $<$2.7(-8) & TW,11,12 \\
83968C & HM & -- & 3.4(3) & 1.9(3) & 14 & 1.23 & 1.37 & 4.9$\pm$0.7(-4) & 1.23 & 1.23 &  1.4$\pm$1.3(17) &  3.0$\pm$2.9(-6) & TW,11,12 \\
83968D & HM & -- & 3.4(3) & 1.8(3) & 14 & 1.23 & 1.37 & 4.7$\pm$0.7(-4) & 1.23 & 1.23 & $<$3.3(15) & $<$6.8(-8) & TW,11,12 \\
83968E & HM & -- & 3.4(3) & 1.7(3) & 14 & 1.23 & 1.37 & 4.6$\pm$0.7(-4) & 1.23 & 1.23 & $<$1.7(15) & $<$3.6(-8) & TW,11,12 \\
86213A & HM & -- & 4.2(3) & 2.0(4) & 39 & 1.23 & 1.37 & 1.5$\pm$0.2(-3) & 1.23 & 1.23 & $<$1.5(15) & $<$4.8(-8) & TW,11,12 \\
86213B & HM & -- & 4.2(3) & 2.0(4) & 38 & 1.23 & 1.37 & 1.4$\pm$0.2(-3) & 1.23 & 1.23 &  3.1$\pm$2.8(16) &  9.9$\pm$9.0(-7) & TW,11,12 \\
86213C & HM & -- & 4.2(3) & 1.1(4) & 20 & 1.23 & 1.37 & 7.5$\pm$1.1(-4) & 1.23 & 1.23 &  8.0$\pm$7.6(16) &  2.6$\pm$2.4(-6) & TW,11,12 \\
101899 & HM & -- & 4.6(3) & 9.2(4) & 1.2(2) & 1.25 & 1.37 & 4.1$\pm$0.6(-3) & 1.25 & 1.25 &  1.1$\pm$0.7(18) &  4.1$\pm$2.9(-5) & TW,11,12 \\
103421 & HM & -- & 4.6(3) & 2.1(4) & 71 & 1.24 & 1.37 & 2.7$\pm$0.4(-3) & 1.24 & 1.24 &  7.0$\pm$6.1(16) &  2.6$\pm$2.3(-6) & TW,11,12 \\
106756A & HM & -- & 4.6(3) & 1.3(5) & 31 & 1.23 & 1.37 & 1.2$\pm$0.2(-3) & 1.23 & 1.23 &  8.4$\pm$7.7(16) &  3.2$\pm$3.0(-6) & TW,11,12 \\
106756B & HM & -- & 4.6(3) & 5.7(4) & 18 & 1.23 & 1.37 & 6.9$\pm$1.0(-4) & 1.23 & 1.23 & $<$4.2(15) & $<$1.6(-7) & TW,11,12 \\
106756C & HM & -- & 4.6(3) & 5.5(4) & 17 & 1.23 & 1.37 & 6.5$\pm$1.0(-4) & 1.23 & 1.23 & $<$5.0(15) & $<$1.9(-7) & TW,11,12 \\
126120A & HM & -- & 4.6(3) & 4.0(3) & 18 & 1.17 & 1.37 & 7.5$\pm$1.1(-4) & 1.17 & 1.17 &  7.7$\pm$7.3(16) &  2.6$\pm$2.5(-6) & TW,11,12 \\
126120B & HM & -- & 4.6(3) & 3.3(3) & 15 & 1.17 & 1.37 & 6.4$\pm$1.0(-4) & 1.17 & 1.17 &  4.3$\pm$4.1(16) &  1.5$\pm$1.4(-6) & TW,11,12 \\
126120C & HM & -- & 4.6(3) & 1.9(3) & 8.8 & 1.17 & 1.37 & 3.7$\pm$0.6(-4) & 1.17 & 1.17 & $<$5.7(15) & $<$2.0(-7) & TW,11,12 \\
126120D & HM & -- & 4.6(3) & 1.5(3) & 7 & 1.17 & 1.37 & 2.9$\pm$0.4(-4) & 1.17 & 1.17 & $<$1.0(15) & $<$3.5(-8) & TW,11,12 \\
126348 & HM & -- & 4.4(3) & 6.8(3) & 1.7(2) & 1.16 & 1.37 & 6.9$\pm$1.0(-3) & 1.16 & 1.16 &  5.9$\pm$3.4(17) &  1.8$\pm$1.1(-5) & TW,11,12 \\
565926A & HM & -- & 4.7(3) & 6.0(3) & 4.4 & 0.58 & 1.37 & 5.4$\pm$0.8(-4) & 0.58 & 0.58 & $<$5.0(15) & $<$4.4(-8) & TW,11,13 \\
565926B & HM & -- & 4.7(3) & 3.3(3) & 2.4 & 0.58 & 1.37 & 2.9$\pm$0.4(-4) & 0.58 & 0.58 & $<$2.5(15) & $<$2.2(-8) & TW,11,13 \\
565926C & HM & -- & 4.7(3) & 3.0(3) & 2.2 & 0.58 & 1.37 & 2.7$\pm$0.4(-4) & 0.58 & 0.58 & $<$2.8(15) & $<$2.4(-8) & TW,11,13 \\
586092A & HM & -- & 1.8(3) & 3.3(3) & 57 & 0.92 & 1.37 & 2.2$\pm$0.3(-3) & 0.92 & 0.92 &  4.3$\pm$1.4(17) &  1.4$\pm$0.5(-6) & TW,11,13 \\
586092B & HM & -- & 1.8(3) & 1.5(3) & 26 & 0.92 & 1.37 & 9.8$\pm$1.5(-4) & 0.92 & 0.92 &  1.5$\pm$1.4(17) &  5.2$\pm$4.9(-7) & TW,11,13 \\
586092C & HM & -- & 1.8(3) & 1.3(3) & 22 & 0.92 & 1.37 & 8.5$\pm$1.3(-4) & 0.92 & 0.92 & $<$2.9(15) & $<$9.7(-9) & TW,11,13 \\
615590 & HM & -- & 2.7(3) & 5.5(3) & 1.2(2) & 0.64 & 1.37 & 9.1$\pm$1.4(-3) & 0.64 & 0.64 &  4.2$\pm$1.1(18) &  1.5$\pm$0.4(-5) & TW,11,13 \\
640076 & HM & -- & 5.5(3) & 2.2(5) & 1.4(2) & 0.87 & 1.37 & 9.9$\pm$1.5(-3) & 0.87 & 0.87 &  1.2$\pm$1.1(17) &  3.2$\pm$3.1(-6) & TW,11,13 \\
644284A & HM & -- & 4.8(3) & -- & 1.9(2) & 0.86 & 1.37 & 1.3$\pm$0.2(-2) & 0.86 & 0.86 &  1.0$\pm$0.5(18) &  2.0$\pm$1.0(-5) & TW,11 \\
644284B & HM & -- & 4.8(3) & -- & 1.1(2) & 0.86 & 1.37 & 7.6$\pm$1.1(-3) & 0.86 & 0.86 &  1.0$\pm$0.5(18) &  2.0$\pm$1.0(-5) & TW,11 \\
693050 & HM & -- & 4.3(3) & 1.2(4) & 1.9(3) & 0.99 & 1.37 & 1.0$\pm$0.2(-1) & 0.99 & 0.99 &  8.9$\pm$4.4(17) &  1.9$\pm$1.0(-5) & TW,11,12 \\
695243 & HM & -- & 4.4(3) & 8.5(2) & 29 & 0.98 & 1.37 & 1.5$\pm$0.2(-3) & 0.98 & 0.98 & $<$2.1(15) & $<$4.6(-8) & TW,11,12 \\
704792 & HM & -- & 3.0(3) & 5.2(4) & 68 & 1.29 & 1.37 & 2.0$\pm$0.3(-3) & 1.29 & 1.29 & $<$1.2(15) & $<$2.2(-8) & TW,11,12 \\
705768 & HM & -- & 6.9(3) & 9.2(4) & 1.4(2) & 0.87 & 1.37 & 1.2$\pm$0.2(-2) & 0.87 & 0.87 &  1.1$\pm$0.7(18) &  4.7$\pm$2.8(-5) & TW,11,12 \\
706733A & HM & -- & 6.8(3) & 1.5(4) & 8.9 & 0.87 & 1.37 & 7.0$\pm$1.1(-4) & 0.87 & 0.87 & $<$4.5(15) & $<$1.8(-7) & TW,11,12 \\
706733B & HM & -- & 6.8(3) & 1.0(4) & 5.9 & 0.87 & 1.37 & 4.7$\pm$0.7(-4) & 0.87 & 0.87 & $<$1.7(15) & $<$6.7(-8) & TW,11,12 \\
706733C & HM & -- & 6.8(3) & 7.6(3) & 4.3 & 0.87 & 1.37 & 3.4$\pm$0.5(-4) & 0.87 & 0.87 & $<$2.4(15) & $<$9.8(-8) & TW,11,12 \\
706785A & HM & -- & 6.4(3) & 5.2(4) & 15 & 0.88 & 1.37 & 1.1$\pm$0.2(-3) & 0.88 & 0.88 &  1.3$\pm$1.2(17) &  4.8$\pm$4.7(-6) & TW,11,12 \\
706785B & HM & -- & 6.4(3) & 4.7(4) & 14 & 0.88 & 1.37 & 1.0$\pm$0.2(-3) & 0.88 & 0.88 &  9.5$\pm$9.4(16) &  3.5$\pm$3.4(-6) & TW,11,12 \\
706785C & HM & -- & 6.4(3) & 3.6(4) & 10 & 0.88 & 1.37 & 7.6$\pm$1.1(-4) & 0.88 & 0.88 & $<$1.5(15) & $<$5.5(-8) & TW,11,12 \\
706785D & HM & -- & 6.4(3) & 2.8(4) & 8 & 0.88 & 1.37 & 6.1$\pm$0.9(-4) & 0.88 & 0.88 & $<$3.9(15) & $<$1.4(-7) & TW,11,12 \\
707948 & HM & -- & 7.1(3) & 2.0(5) & 6.0(2) & 0.88 & 1.37 & 4.8$\pm$0.7(-2) & 0.88 & 0.88 &  1.1$\pm$0.7(19) &  4.9$\pm$3.4(-4) & TW,11,12 \\
717461A & HM & -- & 4.3(3) & 3.3(3) & 55 & 1.29 & 1.37 & 1.9$\pm$0.3(-3) & 1.29 & 1.29 &  7.1$\pm$5.1(17) &  2.6$\pm$1.8(-5) & TW,11,12 \\
717461B & HM & -- & 4.3(3) & 1.4(3) & 23 & 1.29 & 1.37 & 7.9$\pm$1.2(-4) & 1.29 & 1.29 & $<$3.4(15) & $<$1.2(-7) & TW,11,12 \\
721992 & HM & -- & 5.4(3) & 2.6(3) & 47 & 0.85 & 1.37 & 3.5$\pm$0.5(-3) & 0.85 & 0.85 &  2.2$\pm$1.8(17) &  5.5$\pm$4.3(-6) & TW,11,12 \\
724566 & HM & -- & 4.9(3) & 2.3(2) & 89 & 0.83 & 1.37 & 6.5$\pm$1.0(-3) & 0.83 & 0.83 &  1.1$\pm$0.6(18) &  2.2$\pm$1.1(-5) & TW,11,12 \\
732038 & HM & -- & 5.6(3) & 7.5(4) & 72 & 0.82 & 1.37 & 5.6$\pm$0.9(-3) & 0.82 & 0.82 &  1.6$\pm$1.0(17) &  4.0$\pm$2.4(-6) & TW,11,12 \\
744757A & HM & -- & 2.5(3) & 1.2(4) & 1.0(2) & 1.30 & 1.37 & 2.6$\pm$0.4(-3) & 1.30 & 1.30 &  1.2$\pm$0.8(18) &  1.6$\pm$1.0(-5) & TW,11,12 \\
744757B & HM & -- & 2.5(3) & 8.6(3) & 71 & 1.30 & 1.37 & 1.9$\pm$0.3(-3) & 1.30 & 1.30 &  7.5$\pm$6.3(16) &  9.4$\pm$7.9(-7) & TW,11,12 \\
759150A & HM & -- & 3.8(3) & 4.3(3) & 16 & 1.29 & 1.37 & 5.1$\pm$0.8(-4) & 1.29 & 1.29 & $<$2.0(15) & $<$5.6(-8) & TW,11,12 \\
759150B & HM & -- & 3.8(3) & 2.3(3) & 8.1 & 1.29 & 1.37 & 2.7$\pm$0.4(-4) & 1.29 & 1.29 & $<$5.9(14) & $<$1.7(-8) & TW,11,12 \\
759150C & HM & -- & 3.8(3) & 1.5(3) & 5.3 & 1.29 & 1.37 & 1.7$\pm$0.3(-4) & 1.29 & 1.29 & $<$6.2(14) & $<$1.7(-8) & TW,11,12 \\
759150D & HM & -- & 3.8(3) & 1.4(3) & 4.8 & 1.29 & 1.37 & 1.6$\pm$0.2(-4) & 1.29 & 1.29 & $<$1.2(15) & $<$3.4(-8) & TW,11,12 \\
759150E & HM & -- & 3.8(3) & 1.4(3) & 4.8 & 1.29 & 1.37 & 1.6$\pm$0.2(-4) & 1.29 & 1.29 & $<$1.7(15) & $<$4.7(-8) & TW,11,12 \\
767784 & HM & -- & 4.0(3) & 1.4(5) & 7.9(2) & 1.29 & 1.37 & 2.7$\pm$0.4(-2) & 1.29 & 1.29 &  2.1$\pm$0.7(18) &  6.8$\pm$2.3(-5) & TW,11,12 \\
800287 & HM & -- & 4.9(3) & 1.0(5) & 69 & 0.80 & 1.37 & 5.2$\pm$0.8(-3) & 0.80 & 0.80 &  7.8$\pm$2.2(17) &  1.4$\pm$0.4(-5) & TW,11,12 \\
854214A & HM & -- & 4.5(3) & 1.2(4) & 54 & 1.26 & 1.37 & 2.0$\pm$0.3(-3) & 1.26 & 1.26 &  6.4$\pm$5.7(16) &  2.4$\pm$2.2(-6) & TW,11,12 \\
854214B & HM & -- & 4.5(3) & 5.5(3) & 25 & 1.26 & 1.37 & 9.4$\pm$1.4(-4) & 1.26 & 1.26 &  6.2$\pm$6.1(16) &  2.3$\pm$2.2(-6) & TW,11,12 \\
863312A & HM & -- & 4.7(3) & 8.4(4) & 36 & 0.83 & 1.37 & 2.5$\pm$0.4(-3) & 0.83 & 0.83 & $<$3.8(15) & $<$6.9(-8) & TW,11,12 \\
863312B & HM & -- & 4.7(3) & 4.9(4) & 21 & 0.83 & 1.37 & 1.5$\pm$0.2(-3) & 0.83 & 0.83 & $<$2.3(15) & $<$4.2(-8) & TW,11,12 \\
865468A & HM & -- & 3.0(3) & 4.8(4) & 8.2(2) & 1.24 & 1.37 & 2.6$\pm$0.4(-2) & 1.24 & 1.24 &  1.2$\pm$0.2(19) &  1.9$\pm$0.3(-4) & TW,11,12 \\
865468B & HM & -- & 3.0(3) & 1.8(4) & 3.1(2) & 1.24 & 1.37 & 9.7$\pm$1.5(-3) & 1.24 & 1.24 &  4.6$\pm$2.2(17) &  7.7$\pm$3.7(-6) & TW,11,12 \\
865468C & HM & -- & 3.0(3) & 1.3(4) & 2.2(2) & 1.24 & 1.37 & 6.8$\pm$1.0(-3) & 1.24 & 1.24 &  4.6$\pm$2.2(17) &  7.7$\pm$3.7(-6) & TW,11,12 \\
876288 & HM & -- & 6.0(3) & 5.8(4) & 1.7(2) & 0.81 & 1.37 & 1.4$\pm$0.2(-2) & 0.81 & 0.81 &  1.1$\pm$1.0(18) &  3.0$\pm$2.8(-5) & TW,11,12 \\
881427A & HM & -- & 1.5(3) & 9.8(2) & 6.6(2) & 1.23 & 1.37 & 1.5$\pm$0.2(-2) & 1.23 & 1.23 &  5.7$\pm$2.8(18) &  2.3$\pm$1.1(-5) & TW,11,13 \\
881427B & HM & -- & 1.5(3) & 6.5(2) & 4.4(2) & 1.23 & 1.37 & 9.8$\pm$1.5(-3) & 1.23 & 1.23 &  1.5$\pm$0.6(18) &  5.9$\pm$2.5(-6) & TW,11,13 \\
881427C & HM & -- & 1.5(3) & 4.6(2) & 3.1(2) & 1.23 & 1.37 & 6.8$\pm$1.0(-3) & 1.23 & 1.23 &  8.5$\pm$4.2(18) &  3.4$\pm$1.7(-5) & TW,11,13 \\
G023.3891+00.1851 & HM & -- & 1.1(4) & 9.2(4) & 55 & 1.24 & 1.37 & 3.2$\pm$0.5(-3) & 1.24 & 1.24 &  3.7$\pm$1.1(17) &  7.9$\pm$2.3(-5) & TW,11,12 \\
G023.6566-00.1273 & HM & -- & 3.2(3) & 4.8(3) & 85 & 1.24 & 1.37 & 2.7$\pm$0.4(-3) & 1.24 & 1.24 &  7.0$\pm$2.0(17) &  1.3$\pm$0.4(-5) & TW,11,12 \\
G025.6498+01.0491 & HM & -- & 1.2(4) & 4.2(5) & 1.5(2) & 1.17 & 1.37 & 1.1$\pm$0.2(-2) & 1.17 & 1.17 &  2.0$\pm$0.8(18) &  4.7$\pm$1.9(-4) & TW,11,12 \\
G030.1981-00.1691 & HM & -- & 5.8(3) & 2.2(4) & 15 & 0.60 & 1.37 & 1.9$\pm$0.3(-3) & 0.60 & 0.60 &  3.0$\pm$1.5(17) &  4.2$\pm$2.1(-6) & TW,11,12 \\
G233.8306-00.1803 & HM & -- & 3.5(3) & 1.1(4) & 36 & 0.81 & 1.37 & 2.3$\pm$0.3(-3) & 0.81 & 0.81 & $<$2.4(15) & $<$2.2(-8) & TW,11,13 \\
G305.2017+00.2072A1 & HM & -- & 4.0(3) & 2.0(4) & 94 & 1.30 & 1.37 & 3.1$\pm$0.5(-3) & 1.30 & 1.30 &  8.5$\pm$4.3(17) &  2.7$\pm$1.4(-5) & TW,13 \\
G305.2017+00.2072A2 & HM & -- & 4.0(3) & 1.0(4) & 45 & 1.30 & 1.37 & 1.5$\pm$0.2(-3) & 1.30 & 1.30 &  2.4$\pm$2.3(17) &  7.6$\pm$7.4(-6) & TW,13 \\
G310.0135+00.3892 & HM & -- & 3.0(3) & 3.5(4) & 3.7(2) & 1.30 & 1.37 & 1.1$\pm$0.2(-2) & 1.30 & 1.30 &  7.3$\pm$6.8(16) &  1.3$\pm$1.2(-6) & TW,11,12 \\
G314.3197+00.1125 & HM & -- & 8.2(3) & 6.6(4) & 1.0(2) & 1.30 & 1.37 & 4.9$\pm$0.7(-3) & 1.30 & 1.30 &  8.2$\pm$4.1(17) &  1.1$\pm$0.6(-4) & TW,11,12 \\
G316.6412-00.0867 & HM & -- & 2.7(3) & 8.1(3) & 1.4(2) & 1.29 & 1.37 & 3.9$\pm$0.6(-3) & 1.29 & 1.29 &  2.0$\pm$0.8(18) &  3.0$\pm$1.2(-5) & TW,11,12 \\
G318.0489+00.0854B & HM & -- & 3.2(3) & 3.6(4) & 1.5(2) & 1.30 & 1.37 & 4.4$\pm$0.7(-3) & 1.30 & 1.30 &  7.9$\pm$3.9(17) &  1.6$\pm$0.8(-5) & TW,11,12 \\
G318.9480-00.1969A1 & HM & -- & 1.0(4) & 2.1(5) & 2.2(2) & 1.29 & 1.37 & 1.2$\pm$0.2(-2) & 1.29 & 1.29 &  5.5$\pm$3.3(18) &  1.2$\pm$0.7(-3) & TW,11,12 \\
G318.9480-00.1969A2 & HM & -- & 1.0(4) & 6.6(4) & 69 & 1.29 & 1.37 & 3.7$\pm$0.6(-3) & 1.29 & 1.29 &  8.1$\pm$7.7(16) &  1.7$\pm$1.6(-5) & TW,11,12 \\
G323.7399-00.2617B1 & HM & -- & 3.2(3) & 1.7(3) & 1.5(2) & 1.28 & 1.37 & 4.5$\pm$0.7(-3) & 1.28 & 1.28 &  5.7$\pm$3.8(17) &  1.1$\pm$0.7(-5) & TW,13 \\
G323.7399-00.2617B2 & HM & -- & 3.2(3) & 1.5(3) & 1.3(2) & 1.28 & 1.37 & 3.9$\pm$0.6(-3) & 1.28 & 1.28 &  4.7$\pm$2.8(18) &  9.3$\pm$5.6(-5) & TW,13 \\
G323.7399-00.2617B3 & HM & -- & 3.2(3) & 1.2(3) & 1.0(2) & 1.28 & 1.37 & 3.1$\pm$0.5(-3) & 1.28 & 1.28 &  7.4$\pm$6.0(16) &  1.5$\pm$1.2(-6) & TW,13 \\
G323.7399-00.2617B4 & HM & -- & 3.2(3) & 9.5(2) & 78 & 1.28 & 1.37 & 2.4$\pm$0.4(-3) & 1.28 & 1.28 &  7.7$\pm$6.5(16) &  1.5$\pm$1.3(-6) & TW,13 \\
G323.7399-00.2617B5 & HM & -- & 3.2(3) & 7.3(2) & 60 & 1.28 & 1.37 & 1.8$\pm$0.3(-3) & 1.28 & 1.28 &  7.0$\pm$6.9(16) &  1.4$\pm$1.3(-6) & TW,13 \\
G323.7399-00.2617B6 & HM & -- & 3.2(3) & 5.8(2) & 52 & 1.28 & 1.37 & 1.6$\pm$0.2(-3) & 1.28 & 1.28 &  8.9$\pm$8.8(16) &  1.8$\pm$1.7(-6) & TW,13 \\
G323.7399-00.2617B7 & HM & -- & 3.2(3) & 5.8(2) & 54 & 1.28 & 1.37 & 1.6$\pm$0.2(-3) & 1.28 & 1.28 &  4.1$\pm$4.0(16) &  8.2$\pm$8.1(-7) & TW,13 \\
G327.1192+00.5103 & HM & -- & 4.7(3) & 5.4(4) & 70 & 0.81 & 1.37 & 5.2$\pm$0.8(-3) & 0.81 & 0.81 &  1.9$\pm$0.6(18) &  3.3$\pm$1.1(-5) & TW,11,12 \\
G343.1261-00.0623 & HM & -- & 2.0(3) & 3.4(4) & 3.7(2) & 1.25 & 1.37 & 9.3$\pm$1.4(-3) & 1.25 & 1.25 &  4.7$\pm$1.4(17) &  3.4$\pm$1.1(-6) & TW,11,13 \\
G345.5043+00.3480 & HM & -- & 2.0(3) & 2.9(4) & 3.8(2) & 1.25 & 1.37 & 9.5$\pm$1.4(-3) & 1.25 & 1.25 &  5.0$\pm$3.0(18) &  3.7$\pm$2.2(-5) & TW,11,12 \\
G348.7342-01.0359B1 & HM & -- & 2.8(3) & 1.5(4) & 62 & 1.23 & 1.37 & 1.9$\pm$0.3(-3) & 1.23 & 1.23 &  9.6$\pm$8.8(16) &  1.3$\pm$1.2(-6) & TW,13 \\
G348.7342-01.0359B2 & HM & -- & 2.8(3) & 1.0(4) & 43 & 1.23 & 1.37 & 1.3$\pm$0.2(-3) & 1.23 & 1.23 &  1.5$\pm$1.4(17) &  2.1$\pm$2.0(-6) & TW,13 \\
G348.7342-01.0359B3 & HM & -- & 2.8(3) & 9.7(3) & 41 & 1.23 & 1.37 & 1.2$\pm$0.2(-3) & 1.23 & 1.23 & $<$2.1(15) & $<$2.9(-8) & TW,13 \\
\hline
HH~212 & 0 & Y & 400 & 9 & 1.1(2) & 0.35 & 0.85 & 2.7$\pm$0.4(-3) & 0.35 & 0.11 &  1.4$\pm$0.6(18) &  3.2$\pm$1.4(-9) & 14,15,16,17,18 \\
IRAS 16293A & 0 & N & 140 & 18 & 1.0(3) & 0.50 & 0.88 & 1.2$\pm$0.2(-2) & 0.50 & 0.50 &  1.3$\pm$0.4(19) &  7.5$\pm$2.3(-8) & 19,20,21 \\
IRAS 16293B & 0 & N & 140 & 3 & 2.0(3) & 0.50 & 0.88 & 2.5$\pm$0.4(-2) & 0.50 & 0.50 &  1.0$\pm$0.2(19) &  5.9$\pm$1.3(-8) & 19,20,22 \\
L483 & 0 & N & 200 & 12 & 50 & 0.13 & 0.86 & 5.3$\pm$0.8(-3) & 0.13 & 0.34 &  1.7$\pm$0.4(19) &  9.5$\pm$2.1(-8) & 23,24 \\
IRAS~4A1 & 0 & Y & 320 & 9.1 & 3.9(2) & 0.59 & 1.22 & 1.3$\pm$0.2(-2) & 0.59 & 0.22 & $>$1.0(19) & $>$5.8(-8) & TW,1,8,9,25\\
IRAS~4A2 & 0 & N & 320 & 4.5 & 1.9(2) & 0.59 & 1.22 & 6.4$\pm$1.0(-3) & 0.59 & 0.24 & $>$1.0(18) & $>$6.9(-9) & TW,1,8,25 \\
L1527 & 0 & Y & 140 & 2.3 & 1.3(2) & 1.00 & 1.00 & 7.9$\pm$1.2(-4) & 0.50 & 1.00 & $<$2.0(15) & $<$4.6(-11) & 26,27,28,29 \\
BHR~71 & 0 & N & 200 & 11 & 5.9(2) & 0.32 & 0.88 & 1.7$\pm$0.3(-2) & 0.32 & 0.50 &  2.5$\pm$0.5(18) &  2.9$\pm$0.6(-8) & 2,30,31 \\
Serpens SMM1-a & 0 & -- & 436 & 1.1(2) & 8.0(2) & 0.33 & 0.87 & 3.2$\pm$0.5(-2) & 0.33 & 1.20 &  1.1$\pm$0.6(18) &  3.5$\pm$1.9(-7) & 4,27,32,33 \\
B1-bS & 0 & N & 320 & 0.6 & 1.5(2) & 0.39 & 1.22 & 9.2$\pm$1.4(-3) & 0.63 & 0.35 &  3.6$\pm$0.8(17) &  5.3$\pm$1.2(-9) & TW,1,6,34 \\
Ser-emb 1 & 0 & -- & 436 & 4.1 & 1.2(2) & 0.46 & 1.29 & 7.9$\pm$1.2(-3) & 0.46 & 0.35 & $>$8.0(16) & $>$2.2(-9) & 1,5,35 \\
Ser-emb 11W & I & -- & 436 & 4.9 & 70 & 0.53 & 1.29 & 3.7$\pm$0.6(-3) & 0.53 & 0.14 &  3.8$\pm$0.4(18) &  1.6$\pm$0.2(-8) & 1,5,36 \\
B335 & 0 & N & 100 & 0.8 & 4.8 & 0.05 & 1.30 & 4.3$\pm$0.6(-3) & 0.05 & 0.15 & $>$6.7(18) & $>$1.8(-9) & 2,37,38 \\
SVS 13A & Burst & N & 320 & 57 & 1.1(2) & 0.58 & 1.22 & 3.8$\pm$0.6(-3) & 10.00 & 0.30 &  1.4$\pm$0.3(19) &  1.5$\pm$0.3(-7) & TW,1,8,39 \\
L1551 IRS5 & I & Y & 140 & 23 & 1.5(2) & 0.19 & 1.30 & 2.1$\pm$0.3(-2) & 0.19 & 0.15 & $>$2.0(19) & $>$1.0(-8) & 2,26,40,41 \\
GSS30-IRS1 & I & Y & 139 & 11 & 30 & 0.37 & 0.88 & 5.8$\pm$0.9(-4) & 0.37 & 0.15 & $<$2.5(15) & $<$1.3(-12) & 42,43,44 \\
$[$GY92$]$ 30 & I & Y & 139 & 0.1 & 80 & 0.37 & 0.88 & 1.6$\pm$0.2(-3) & 0.37 & 0.17 & $<$2.5(15) & $<$1.6(-12)) & 42,43,44 \\
WL 12 & I & Y & 139 & 1.4 & 1.4(2) & 0.37 & 0.88 & 2.8$\pm$0.4(-3) & 0.37 & 0.12 & $<$2.5(15) & $<$8.2(-13)) & 42,43,44 \\
$[$GY92$]$ 197 & I & Y & 139 & 0.2 & 71 & 0.37 & 0.88 & 1.4$\pm$0.2(-3) & 0.37 & 0.24 & $<$2.5(15) & $<$3.3(-12)) & 42,43,44 \\
Elias 29 & I & Y & 139 & 18 & 34 & 0.37 & 0.88 & 6.6$\pm$1.0(-4) & 0.37 & 0.16 & $<$2.5(15) & $<$1.5(-12)) & 42,43,44 \\
IRS 43 & I & Y & 139 & 3.3 & 32 & 0.37 & 0.88 & 6.3$\pm$0.9(-4) & 0.37 & 0.17 & $<$2.5(15) & $<$1.6(-12)) & 42,43,44 \\
IRS 44 & I & Y & 139 & 7.1 & 29 & 0.37 & 0.88 & 5.7$\pm$0.9(-4) & 0.37 & 0.21 & $<$2.5(15) & $<$2.5(-12)) & 42,43,44 \\
IRAS 16253 & 0 & Y & 139 & 0.2 & 32 & 0.37 & 0.88 & 6.2$\pm$0.9(-4) & 0.37 & 0.17 & $<$2.5(15) & $<$1.6(-12)) & 42,43,44 \\
ISO-Oph 203 & I & Y & 139 & 0.1 & 11 & 0.37 & 0.88 & 2.1$\pm$0.3(-4) & 0.37 & 0.08 & $<$2.5(15) & $<$3.6(-13)) & 42,43,44 \\
IRS 67 & I & Y & 139 & 2.8 & 1.3(2) & 0.37 & 0.88 & 2.4$\pm$0.4(-3) & 0.37 & 0.17 & $<$2.5(15) & $<$1.6(-12)) & 42,43,44 \\
V883 Ori & Burst & Y & 400 & 2.2(2) & 2.9(2) & 0.20 & 0.88 & 2.4$\pm$0.4(-2) & 0.20 & 0.60 &  4.3$\pm$0.7(17) &  2.9$\pm$0.5(-8) & 14,45,46,47\\
AFGL 4176 & HM & Y & 3.7(3) & 9.7(4) & 29 & 0.30 & 1.20 & 5.3$\pm$0.8(-3) & 0.30 & 0.30 &  5.5$\pm$0.4(18) &  8.0$\pm$0.6(-6) & 48,49,50,51 \\
Sgr B2(N2) & HM & -- & 8.3(3) & 2.6(5) & 1.4(2) & 1.50 & 3.00 & 8.2$\pm$1.2(-2) & 1.50 & 1.40 &  4.1$\pm$0.9(19) &  6.5$\pm$1.4(-3) & 52,53,54,55 \\
Sgr B2(N3) & HM & -- & 8.3(3) & 4.5(4) & -- & -- & -- & -- & 1.50 & 0.40 &  4.1$\pm$0.9(18) &  5.3$\pm$1.2(-5) & 52,53,54,56 \\
Sgr B2(N4) & HM & -- & 8.3(3) & 3.9(5) & 34 & 1.50 & 3.00 & 2.0$\pm$0.3(-2) & 1.50 & 1.00 &  2.6$\pm$0.6(17) &  2.1$\pm$0.5(-5) & 52,53,54,56 \\
Sgr B2(N5) & HM & -- & 8.3(3) & 2.8(5) & 62 & 1.50 & 3.00 & 3.7$\pm$0.6(-2) & 1.50 & 1.00 &  9.7$\pm$2.1(17) &  7.9$\pm$1.7(-5) & 52,53,54,56 \\
NGC6334I MM1 I & HM & -- & 1.3(3) & 5.8(4) & 2.4(3) & 0.53 & 1.30 & 1.4$\pm$0.2(-1) & 0.20 & 0.87 &  1.2$\pm$0.3(20) &  1.8$\pm$0.5(-4) & 52,57,58,59 \\
NGC6334I MM1 II & HM & -- & 1.3(3) & 1.1(4) & 4.5(2) & 0.54 & 1.30 & 2.7$\pm$0.4(-2) & 0.20 & 0.87 &  9.9$\pm$0.9(19) &  1.5$\pm$0.1(-4) & 52,57,58,59 \\
NGC6334I MM1 III & HM & -- & 1.3(3) & 5.1(4) & 2.1(3) & 0.50 & 1.30 & 1.4$\pm$0.2(-1) & 0.20 & 0.87 &  7.8$\pm$1.7(19) &  1.2$\pm$0.3(-4) & 52,57,58,59 \\
NGC6334I MM1 IV & HM & -- & 1.3(3) & 1.1(4) & 4.5(2) & 0.54 & 1.30 & 2.7$\pm$0.4(-2) & 0.20 & 0.87 &  6.9$\pm$1.5(19) &  1.0$\pm$0.2(-4) & 52,57,58,59 \\
NGC6334I MM1 V & HM & -- & 1.3(3) & 1.1(4) & 4.5(2) & 0.55 & 1.30 & 2.6$\pm$0.4(-2) & 0.20 & 0.87 &  1.9$\pm$0.4(19) &  2.9$\pm$0.6(-5) & 52,57,58,59 \\
NGC6334I MM2 I & HM & -- & 1.3(3) & 7.5(3) & 3.2(2) & 0.34 & 1.30 & 3.8$\pm$0.6(-2) & 0.20 & 0.87 &  5.2$\pm$1.6(19) &  7.8$\pm$2.4(-5) & 52,57,58,59 \\
NGC6334I MM2 II & HM & -- & 1.3(3) & 1.5(3) & 8 & 0.18 & 1.30 & 2.5$\pm$0.4(-3) & 0.20 & 0.87 &  1.5$\pm$0.3(19) &  2.2$\pm$0.5(-5) & 52,57,58,59 \\
NGC6334I MM3 I & HM & -- & 1.3(3) & -- & -- & -- & -- & -- & 0.20 & 0.87 &  6.5$\pm$1.4(18) &  9.7$\pm$2.1(-6) & 52,59 \\
NGC6334I MM3 II & HM & -- & 1.3(3) & -- & -- & -- & -- & -- & 0.20 & 0.87 &  6.0$\pm$1.3(18) &  9.0$\pm$2.0(-6) & 52,59 \\
\end{longtable}
\tablefoot{$a(b)$ represents $a\times10^b$. In the disk column we indicate whether a disk is confirmed (Y), whether no disk on $>50$~au scales is confirmed (N), or whether no information about a disk is available (--). \\ \\
{\bf References. } References for distances, bolometric luminosities (scaled to the reported distance), continuum fluxes, column densities, and, if applicable, a disk detection:
TW: This work; 
1: \citet{Ortiz-Leon2018};
2: \citet{Karska2018};
3: \citet{vanGelder2020};
4: \citet{Ortiz-Leon2017};
5: \citet{Enoch2011};
6: \citet{Murillo2016};
7: \citet{Enoch2009};
8: \citet{Tobin2016};
9: \citet{Segura-Cox2018};
10: \citet{Maury2019};
11: \citet{Mege2021};
12: \citet{Elia2017};
13: \citet{Lumsden2013};
14: \citet{Kounkel2017};
15: \citet{Zinnecker1992};
16: \citet{Lee2008};
17: \citet{Lee2019_HH212};
18: \citet{Lee2017_HH212_disk};
19: \citet{Dzib2018};
20: \citet{Jacobsen2018};
21: \citet{Manigand2020};
22: \citet{Jorgensen2018};
23: \citet{Jacobsen2019};
24: \citet{Tafalla2000};
25: \citet{deSimone2020};
26: \citet{Zucker2019};
27: \citet{Kristensen2012};
28: \citet{Belloche2020};
29: \citet{Tobin2012};
30: \citet{Seidensticker1989};
31: \citet{Yang2020};
32: \citet{Hull2017};
33: \citet{Ligterink2021};
34: \citet{Marcelino2018};
35: \citet{Martin-domenech2019};
36: \citet{Martin-domenech2021};
37: \citet{Olofsson2009};
38: \citet{Bjerkeli2019};
39: \citet{Bianchi2017};
40: \citet{Bianchi2020};
41: \citet{Cruz-SaenzdeMiera2019};
42: \citet{Mamajek2008};
43: \citet{Dunham2015};
44: \citet{ArturdelaVillarmois2019};
45: \citet{Furlan2016};
46: \citet{Lee_V883_2019};
47: \citet{Cieza2016};
48: \citet{Bailer-Jones2018};
49: \citet{Beltran2006};
50: \citet{Bogelund2019};
51: \citet{Johnston2015};
52: \citet{Reid2014};
53: \citet{Bonfand2019};
54: \citet{Belloche2016};
55: \citet{Muller2016};
56: \citet{Bonfand2017};
57: \citet{Sandell2000};
58: \citet{Brogan2016};
59: \citet{Bogelund2018}.
}
\end{landscape}

\section{Transitions of CH$_3$OH and isotopologues}
\label{app:CH3OH_transitions}
\begin{longtable}{lrclccl}
\caption{Transitions of CH$_3$OH and isotopologues covered in the various ALMA programs. \label{tab:CH3OH_transitions}} \\
\hline\hline
Species & \multicolumn{3}{c}{Transition} & Frequency & $A_\mathrm{ij}$ & $E_\mathrm{up}$ \\
 & (J K L M & - & J K L M) & (GHz) & (s$^{-1}$) & (K) \\
\hline
\endfirsthead
\caption{continued.}\\
\hline\hline
Species & \multicolumn{3}{c}{Transition} & Frequency & $A_\mathrm{ij}$ & $E_\mathrm{up}$ \\
 & (J K L M & - & J K L M) & (GHz) & (s$^{-1}$) & (K) \\
\hline
\endhead
\hline
\endfoot
\multicolumn{7}{c}{2017.1.01174.S ($\theta_\mathrm{beam}\sim0.45^{\prime\prime}$, $\mathrm{rms_{line}}\sim0.15$~K)} \\
\hline
CH$_3$OH & 20 8 13 2 & - & 21 7 14 2 & 260.0643 & $2.1 \times 10^{-5}$ & 808.2 \\
 & 20 3 18 0 & - & 20 2 19 0 & 260.3815 & $9.1 \times 10^{-5}$ & 536.9 \\
 & 2 1 1 1 & - & 1 0 1 1 & 261.8057 & $5.6 \times 10^{-5}$ & 28.0 \\
\hline
$^{13}$CH$_3$OH & 2 1 1 0 & - & 1 0 1 0 & 259.9865 & $5.5 \times 10^{-5}$ & 27.9 \\
 & 18 3 16 +0 & - & 18 2 17 -0 & 260.0888 & $9.2 \times 10^{-5}$ & 437.2 \\
 & 20 3 18 +0 & - & 20 2 19 -0 & 262.7673 & $9.4 \times 10^{-5}$ & 525.4 \\
 & 7 4 4 -0 & - & 8 3 5 -0 & 262.9130 & $1.7 \times 10^{-5}$ & 144.2 \\
 & 7 4 3 +0 & - & 8 3 6 +0 & 262.9203 & $1.7 \times 10^{-5}$ & 144.2 \\
 & 5 2 3 0 & - & 4 1 3 0 & 263.1133 & $7.4 \times 10^{-5}$ & 56.3 \\
 & 11 2 10 -0 & - & 10 3 7 -0 & 263.3060 & $2.9 \times 10^{-5}$ & 187.3 \\
 & 25 -7 18 0 & - & 26 -6 20 0 & 263.5779 & $2.8 \times 10^{-5}$ & 996.5 \\
\hline
CH$_3^{18}$OH & 12 2 10 4 & - & 13 3 10 4 & 260.0355 & $7.1 \times 10^{-5}$ & 537.2 \\
 & 16 3 13 0 & - & 16 2 14 0 & 260.3163 & $9.3 \times 10^{-5}$ & 352.6 \\
 & 15 1 15 1 & - & 14 2 13 1 & 260.4403 & $2.8 \times 10^{-5}$ & 272.2 \\
 & 12 3 9 0 & - & 12 2 10 0 & 261.8972 & $9.2 \times 10^{-5}$ & 223.6 \\
 & 11 3 8 0 & - & 11 2 9 0 & 262.1524 & $9.2 \times 10^{-5}$ & 196.9 \\
 & 7 3 4 0 & - & 7 2 5 0 & 262.7722 & $8.6 \times 10^{-5}$ & 112.4 \\
 & 6 3 3 0 & - & 6 2 4 0 & 262.8532 & $8.3 \times 10^{-5}$ & 96.8 \\
 & 5 3 2 0 & - & 5 2 3 0 & 262.9139 & $7.7 \times 10^{-5}$ & 83.5 \\
 & 4 3 1 0 & - & 4 2 2 0 & 262.9583 & $6.8 \times 10^{-5}$ & 72.3 \\
 & 5 3 3 0 & - & 5 2 4 0 & 262.9825 & $7.7 \times 10^{-5}$ & 83.5 \\
 & 4 3 2 0 & - & 4 2 3 0 & 262.9878 & $6.8 \times 10^{-5}$ & 72.3 \\
 & 3 3 0 0 & - & 3 2 1 0 & 262.9897 & $4.8 \times 10^{-5}$ & 63.4 \\
 & 6 3 4 0 & - & 6 2 5 0 & 262.9900 & $8.3 \times 10^{-5}$ & 96.8 \\
 & 3 3 1 0 & - & 3 2 2 0 & 262.9995 & $4.8 \times 10^{-5}$ & 63.4 \\
 & 7 3 5 0 & - & 7 2 6 0 & 263.0177 & $8.6 \times 10^{-5}$ & 112.4 \\
 & 8 3 6 0 & - & 8 2 7 0 & 263.0739 & $8.9 \times 10^{-5}$ & 130.2 \\
 & 9 3 7 0 & - & 9 2 8 0 & 263.1684 & $9.0 \times 10^{-5}$ & 150.2 \\
 & 10 3 8 0 & - & 10 2 9 0 & 263.3116 & $9.2 \times 10^{-5}$ & 172.5 \\
 & 16 5 12 4 & - & 17 6 12 4 & 263.4603 & $5.8 \times 10^{-5}$ & 846.9 \\
 & 11 3 9 0 & - & 11 2 10 0 & 263.5151 & $9.2 \times 10^{-5}$ & 196.9 \\
\hline
\multicolumn{7}{c}{2017.1.01350.S ($\theta_\mathrm{beam}\sim0.3^{\prime\prime}$, $\mathrm{rms_{line}}\sim0.5$~K)} \\
\hline
CH$_3$OH & 20 1 19 1 & - & 20 0 20 1 & 217.8865 & $3.4 \times 10^{-5}$ & 508.4 \\
 & 25 3 23 1 & - & 24 4 20 1 & 219.9837 & $2.0 \times 10^{-5}$ & 802.2 \\
 & 23 5 18 1 & - & 22 6 17 1 & 219.9937 & $1.7 \times 10^{-5}$ & 775.9 \\
 & 10 5 6 2 & - & 11 4 8 2 & 220.4013 & $1.1 \times 10^{-5}$ & 251.6 \\
 & 10 2 9 0 & - & 9 3 6 0 & 231.2811 & $1.8 \times 10^{-5}$ & 165.3 \\
 & 11 6 5 4 & - & 11 7 4 4 & 233.1212 & $1.0 \times 10^{-5}$ & 745.0 \\
 & 18 3 15 0 & - & 17 4 14 0 & 233.7957 & $2.2 \times 10^{-5}$ & 446.6 \\
 & 13 3 11 6 & - & 14 4 11 6 & 233.9170 & $1.4 \times 10^{-5}$ & 868.5 \\
 & 13 3 10 6 & - & 14 4 10 6 & 233.9170 & $1.4 \times 10^{-5}$ & 868.5 \\
 & 4 2 3 0 & - & 5 1 4 0 & 234.6834 & $1.9 \times 10^{-5}$ & 60.9 \\
\hline
$^{13}$CH$_3$OH & 14 1 13 -0 & - & 13 2 12 -0 & 217.0446 & $2.4 \times 10^{-5}$ & 254.3 \\
 & 21 -4 18 0 & - & 20 -5 15 0 & 231.3891 & $2.1 \times 10^{-5}$ & 611.1 \\
 & 24 5 20 0 & - & 23 6 18 0 & 233.4879 & $2.2 \times 10^{-5}$ & 815.4 \\
 & 5 1 5 +0 & - & 4 1 4 +0 & 234.0116 & $5.3 \times 10^{-5}$ & 48.3 \\
 & 9 1 9 1 & - & 8 0 8 1 & 234.5609 & $3.6 \times 10^{-5}$ & 393.3 \\
 & 26 3 23 0 & - & 25 4 21 0 & 234.8658 & $2.5 \times 10^{-5}$ & 843.0 \\
\hline
CH$_3^{18}$OH & 26 3 23 2 & - & 25 4 21 2 & 216.6903 & $1.9 \times 10^{-5}$ & 828.8 \\
 & 14 1 14 1 & - & 13 2 12 1 & 217.1729 & $1.7 \times 10^{-5}$ & 238.9 \\
 & 18 6 13 4 & - & 17 7 11 4 & 217.9223 & $1.5 \times 10^{-5}$ & 874.1 \\
 & 5 1 5 6 & - & 4 1 4 6 & 230.4823 & $5.0 \times 10^{-5}$ & 712.0 \\
 & 11 0 11 0 & - & 10 1 10 0 & 230.9519 & $6.4 \times 10^{-5}$ & 146.8 \\
 & 5 2 3 8 & - & 4 2 2 8 & 230.9544 & $4.4 \times 10^{-5}$ & 617.8 \\
 & 5 1 5 7 & - & 4 1 4 7 & 230.9582 & $5.1 \times 10^{-5}$ & 563.6 \\
 & 5 1 5 3 & - & 4 1 4 3 & 230.9590 & $5.1 \times 10^{-5}$ & 358.0 \\
 & 5 4 2 5 & - & 4 4 1 5 & 231.1535 & $1.9 \times 10^{-5}$ & 397.3 \\
 & 5 3 3 5 & - & 4 3 2 5 & 231.1639 & $3.4 \times 10^{-5}$ & 450.3 \\
 & 5 4 1 4 & - & 4 4 0 4 & 231.1689 & $1.9 \times 10^{-5}$ & 437.7 \\
 & 5 3 2 4 & - & 4 3 1 4 & 231.1709 & $3.4 \times 10^{-5}$ & 356.1 \\
 & 5 4 1 3 & - & 4 4 0 3 & 231.1757 & $1.9 \times 10^{-5}$ & 514.0 \\
 & 5 4 2 3 & - & 4 4 0 3 & 231.1757 & $1.9 \times 10^{-5}$ & 514.0 \\
 & 5 2 3 4 & - & 4 2 2 4 & 231.1787 & $4.4 \times 10^{-5}$ & 397.4 \\
 & 5 2 3 3 & - & 4 2 2 3 & 231.1820 & $4.4 \times 10^{-5}$ & 331.8 \\
 & 5 3 3 3 & - & 4 3 2 3 & 231.1836 & $3.4 \times 10^{-5}$ & 428.0 \\
 & 5 3 2 3 & - & 4 3 1 3 & 231.1836 & $3.4 \times 10^{-5}$ & 428.0 \\
 & 5 2 4 3 & - & 4 2 3 3 & 231.1852 & $4.4 \times 10^{-5}$ & 331.8 \\
 & 5 1 5 5 & - & 4 1 4 5 & 231.1917 & $5.1 \times 10^{-5}$ & 324.4 \\
 & 5 0 5 4 & - & 4 0 4 4 & 231.1942 & $5.3 \times 10^{-5}$ & 333.4 \\
 & 5 2 4 5 & - & 4 2 3 5 & 231.1962 & $4.5 \times 10^{-5}$ & 431.4 \\
 & 5 1 4 4 & - & 4 1 3 4 & 231.2260 & $5.1 \times 10^{-5}$ & 444.6 \\
 & 5 0 5 3 & - & 4 0 4 3 & 231.2553 & $5.3 \times 10^{-5}$ & 454.7 \\
 & 5 1 4 6 & - & 4 1 3 6 & 231.3405 & $5.1 \times 10^{-5}$ & 712.1 \\
 & 17 2 16 3 & - & 16 1 15 3 & 233.6336 & $4.0 \times 10^{-5}$ & 637.9 \\
 & 5 1 4 0 & - & 4 1 3 0 & 233.7279 & $5.3 \times 10^{-5}$ & 48.0 \\
\hline
\multicolumn{7}{c}{2016.1.01501.S \& 2017.1.01426.S ($\theta_\mathrm{beam}\sim0.4^{\prime\prime}$, $\mathrm{rms_{line}}\sim0.2$~K)} \\
\hline
CH$_3$OH & 5 1 4 0 & - & 4 1 3 0 & 243.9158 & $6.0 \times 10^{-5}$ & 49.7 \\
 & 20 3 17 0 & - & 20 2 18 0 & 246.0746 & $8.2 \times 10^{-5}$ & 537.0 \\
 & 19 3 16 0 & - & 19 2 17 0 & 246.8733 & $8.3 \times 10^{-5}$ & 490.7 \\
 & 2 1 1 1 & - & 1 0 1 1 & 261.8057 & $5.6 \times 10^{-5}$ & 28.0 \\
\hline
$^{13}$CH$_3$OH & 23 4 19 0 & - & 22 5 18 0 & 246.4261 & $2.6 \times 10^{-5}$ & 721.0 \\
 & 23 3 20 -0 & - & 23 2 21 +0 & 247.0863 & $8.5 \times 10^{-5}$ & 674.9 \\
 & 17 3 15 +0 & - & 17 2 16 -0 & 259.0365 & $9.1 \times 10^{-5}$ & 396.5 \\
\hline
CH$_3^{18}$OH & 11 2 10 0 & - & 10 3 7 0 & 246.2566 & $2.3 \times 10^{-5}$ & 184.3 \\
 & 15 1 14 4 & - & 16 2 14 4 & 246.8637 & $1.3 \times 10^{-5}$ & 677.5 \\
\hline
\multicolumn{7}{c}{2019.1.00195.L ($\theta_\mathrm{beam}\sim1^{\prime\prime}$, $\mathrm{rms_{line}}\sim0.2$~K)} \\
\hline
CH$_3$OH & 6 1 5 3 & - & 7 2 5 3 & 217.2992 & $4.3 \times 10^{-5}$ & 373.9 \\
 & 15 6 9 3 & - & 16 5 11 3 & 217.6427 & $1.9 \times 10^{-5}$ & 745.6 \\
 & 15 6 10 3 & - & 16 5 12 3 & 217.6427 & $1.9 \times 10^{-5}$ & 745.6 \\
 & 20 1 19 1 & - & 20 0 20 1 & 217.8865 & $3.4 \times 10^{-5}$ & 508.4 \\
 & 4 2 3 1 & - & 3 1 2 1 & 218.4401 & $4.7 \times 10^{-5}$ & 45.5 \\
 & 25 3 23 1 & - & 24 4 20 1 & 219.9837 & $2.0 \times 10^{-5}$ & 802.2 \\
 & 23 5 18 1 & - & 22 6 17 1 & 219.9937 & $1.7 \times 10^{-5}$ & 775.9 \\
 & 8 0 8 1 & - & 7 1 6 1 & 220.0786 & $2.5 \times 10^{-5}$ & 96.6 \\
 & 10 5 6 2 & - & 11 4 8 2 & 220.4013 & $1.1 \times 10^{-5}$ & 251.6 \\
\hline
$^{13}$CH$_3$OH & 14 1 13 -0 & - & 13 2 12 -0 & 217.0446 & $2.4 \times 10^{-5}$ & 254.3 \\
 & 10 2 8 +0 & - & 9 3 7 +0 & 217.3995 & $1.5 \times 10^{-5}$ & 162.4 \\
 & 17 7 11 +0 & - & 18 6 12 +0 & 220.3218 & $1.3 \times 10^{-5}$ & 592.3 \\
 & 17 7 10 -0 & - & 18 6 13 -0 & 220.3218 & $1.3 \times 10^{-5}$ & 592.3 \\
\hline
CH$_3^{18}$OH & 14 1 14 1 & - & 13 2 12 1 & 217.1729 & $1.7 \times 10^{-5}$ & 238.9 \\
 & 18 6 13 4 & - & 17 7 11 4 & 217.9223 & $1.5 \times 10^{-5}$ & 874.1 \\
 & 17 5 13 4 & - & 18 6 13 4 & 218.5521 & $3.2 \times 10^{-5}$ & 884.6 \\
 & 21 1 20 2 & - & 21 0 21 1 & 218.8351 & $3.4 \times 10^{-5}$ & 534.5 \\
 & 23 4 19 2 & - & 22 5 18 2 & 219.0356 & $1.8 \times 10^{-5}$ & 709.6 \\
 & 4 2 2 2 & - & 3 1 2 2 & 219.4078 & $4.6 \times 10^{-5}$ & 44.6 \\
 & 8 7 1 5 & - & 7 6 1 5 & 219.8433 & $2.8 \times 10^{-5}$ & 663.2 \\
 & 18 3 16 5 & - & 19 4 16 5 & 219.9572 & $5.1 \times 10^{-5}$ & 795.8 \\
 & 8 1 8 1 & - & 7 0 7 1 & 220.1951 & $3.6 \times 10^{-5}$ & 85.7 \\
\end{longtable}
\tablefoot{The typical beam size and rms are listed for each dataset.}

\section{Toy model of a spherically symmetric infalling envelope}
\label{app:toy_model}
In this simple toy model of a spherically symmetric infalling envelope, the envelope is heated by the luminosity of the central protostar. The density structure at a radius $R$ can be written as,
\begin{align}
n_\mathrm{H} = n_\mathrm{H,0} \left(\frac{R}{R_\mathrm{0}}\right)^{-p},
\label{eq:density_structure}
\end{align}
with $n_\mathrm{H,0}$ the density at a radius $R_\mathrm{0}$. The snowline radius of a molecule with sublimation temperature $T_\mathrm{sub}$ then scales as \citep[for details, see Appendix~B of][]{Nazari2021}, 
\begin{align}
R_\mathrm{sub} = R_\mathrm{0} \left(\frac{T_\mathrm{sub}}{T_\mathrm{0}}\right)^{-1/\gamma},
\label{eq:snowline_radius}
\end{align}
where $\gamma$ is the slope of the temperature profile in the envelope and $T_\mathrm{0}$ is the temperature at a radius $R_\mathrm{0}$. In this work, a sublimation temperature of $100$~K is adopted based on experiments of pure CH$_3$OH and mixed H$_2$O:CH$_3$OH ices \citep[e.g.,][]{Collings2004}. At a typical inner envelope density of $10^{7}$~cm$^{-3}$, the adopted $T_\mathrm{sub}$ corresponds to a binding energy of $\sim5000$~K which is slightly higher than that recommended by \citet{Penteado2017}. Radiative transfer calculations show that for protostellar envelopes $\gamma = 2/5$ at $>10$~au scales and that $T_\mathrm{0} \propto L_\mathrm{bol}^{1/5}$ \citep{Adams1985}. Eq.~\eqref{eq:snowline_radius} can then be written as,
\begin{align}
R_\mathrm{sub} \propto L_\mathrm{bol}^{1/2},
\label{eq:snowline_Lbol_relation}
\end{align}
which (except for a prefactor) is similar to the relation that was derived by \citet{Bisschop2007} for the 100~K radius in high-mass protostars and more recently seen to be remarkably similiar for low-mass protostars \citep{vantHoff2022}. Using Eqs.~\eqref{eq:density_structure} and \eqref{eq:snowline_radius}, the total warm methanol mass inside the snowline is proportional to \citep{Nazari2021},
\begin{align}
M_\mathrm{CH_3OH} \propto n_\mathrm{H,0} L_\mathrm{bol}^{(3-p)/(5\gamma)}.
\end{align}
Filling in that $p=3/2$ for an infalling envelope, $\gamma = 2/5$, and assuming that $n_\mathrm{H,0}\propto M_\mathrm{0}$ (i.e., a more massive envelope is denser at a certain radius $R_\mathrm{0}$) yields,
\begin{align}
M_\mathrm{CH_3OH} \propto M_\mathrm{0} L_\mathrm{bol}^{3/4},
\end{align}
where $M_\mathrm{0}$ is the total warm + cold mass contained within a radius $R_\mathrm{0}$. 

\section{Calculating the reference dust mass}
\label{app:M_dust,0}
The dust mass is estimated from the peak continuum flux $F_\mathrm{cont}$ within the central beam. Since the continuum fluxes in this work are derived from multiple datasets covering different frequencies (but all in Band~6), the measured fluxes are scaled to a wavelength of 1.2~mm (i.e., $\sim250$~GHz) through, 
\begin{align}
F_\mathrm{1.2mm} = F_\mathrm{cont} \left(\frac{\lambda_\mathrm{cont}}{1.2~\mathrm{mm}}\right)^\alpha,
\label{eq:continuum_scaling}
\end{align}
where $\lambda_\mathrm{cont}$ is the wavelength of the observations and $\alpha$ is the power-law index of the dust continuum. In this work, $\alpha=2.5$ is assumed for the low-mass sources \citep{Tychoniec2020} and $\alpha=3.5$ for the high-mass sources \citep{Palau2014}. The dust mass within the beam, $M_\mathrm{dust,beam}$ is calculated from the continuum flux using the equation from \citet{Hildebrand1983},
\begin{align}
M_\mathrm{dust,beam} = \frac{F_\mathrm{1.2mm} d^2}{\kappa_\mathrm{\nu} B_\mathrm{\nu}(T_\mathrm{dust})},
\label{eq:M_dust,beam}
\end{align}
with $B_\mathrm{\nu}$ the Planck function for a given dust temperature $T_\mathrm{dust}$ and $\kappa_\mathrm{\nu}$ the dust opacity in the optically thin limit. Here, a dust temperature of 30~K is assumed for low-mass ($L_\mathrm{bol}\lesssim100$~L$_\odot$) sources which is typical for inner envelopes of embedded protostellar systems \citep{Whitney2003}. For high-mass sources ($L_\mathrm{bol}\gtrsim1000$~L$_\odot$), a dust temperature of $50$~K is adopted. Although the dust temperature in the inner regions may be as high as (a few) 100~K, the dust emitting at millimeter wavelengths in our $\sim1^{\prime\prime}$ beam likely resides at larger distances ($\sim1000$~au) from the source where the dust temperature is on the order of 50~K \citep[e.g.,][]{vanderTak2013,Palau2014}. The dust opacity $\kappa_\mathrm{\nu}$ is set to $2.3$~cm$^2$~g$^{-1}$ for a wavelength of 1.2~mm \citep[e.g.,][]{Ansdell2016}. 

$M_\mathrm{dust,beam}$ is an estimate of the dust mass within the beam with a physical radius $R_\mathrm{beam}$ (see Eq.~\eqref{eq:R_source}) \citep[e.g.,][]{Saraceno1996}. The observational data used in this work probe different physical scales depending on the beam size and distance to the source. For example, the PEACHES beam of $\sim0.4-0.5^{\prime\prime}$ probes the disk and inner envelope on $R_\mathrm{beam}\sim50$~au scales in Perseus while the ALMAGAL beam of $\sim1-2^{\prime\prime}$ probes a larger fraction of the envelope at $R_\mathrm{beam}\sim1000-5000$~au scales for sources at $\sim1-5$~kpc. Therefore, the dust masses are scaled to a common arbitrary radius of $R_0 = 200$~au assuming the mass contained in the beam follows a typical free-falling envelope power-law scaling of $p=3/2$ (see Eq.~\eqref{eq:density_structure}),
\begin{align}
M_\mathrm{dust,0} = M_\mathrm{dust,beam} \left(\frac{R_0}{R_\mathrm{beam}}\right)^{3-p}.
\label{eq:M_dust,0}
\end{align}
The derived values of $M_\mathrm{dust,0}$ are listed in Table~\ref{tab:properties_quantities} for all sources. The uncertainty in $M_\mathrm{dust,0}$ is estimated based on a 10~\% uncertainty on the observed continuum flux $F_\mathrm{cont}$ and a 50~\% uncertainty in the assumed $T_\mathrm{dust}$ for both the low-mass and high-mass sources. Spatial sampling in the PEACHES observations (i.e., resolving out large-scale continuum emission on $\gtrsim500$~au scales) compared to the ALMAGAL observations does not affect our estimated dust masses here since only the peak continuum flux within the central beam is used for the dust mass estimate.

\end{appendix}

\end{document}